\documentclass[aps, prd, preprint, amsfonts,
  amssymb, amsmath, showpacs, a4paper,nofootinbib,superscriptaddress,raggedbottom]{revtex4-1}
  
\usepackage[T1]{fontenc}

\usepackage[utf8]{inputenc}

\usepackage{slashed}
\usepackage{braket}
\usepackage{bm}

\newcommand{\be}{\begin{equation}}
\newcommand{\ee}{\end{equation}}
\newcommand{\bea}{\begin{eqnarray}}
\newcommand{\eea}{\end{eqnarray}}

\usepackage{setspace}

\usepackage{enumitem}

\newlist{listcases}{itemize}{1}
\setlist[listcases,1]{label={},leftmargin=0pt}

\usepackage{color}

\begin{document} 

\count\footins=1000

\title{Discrete spacetime symmetries and particle mixing in non-Hermitian 
scalar quantum field theories}

\author{\vspace{-2mm}Jean Alexandre}
\email{jean.alexandre@kcl.ac.uk}
\affiliation{Department of Physics, King's College London,\\
London WC2R 2LS, United Kingdom}

\author{John Ellis}
\email{john.ellis@cern.ch}
\affiliation{Department of Physics, King's College London,\\ 
London WC2R 2LS, United Kingdom}
\affiliation{National Institute of Chemical Physics \& Biophysics, R\"avala 10, 10143 Tallinn, Estonia}
\affiliation{Theoretical Physics Department, CERN, CH-1211 Geneva 23, Switzerland}

\author{Peter Millington}
\email{p.millington@nottingham.ac.uk}
\affiliation{School of Physics and Astronomy, University of Nottingham,\\ Nottingham NG7 2RD, United Kingdom\vspace{-1mm}}

\begin{abstract}
{\small
We discuss {second quantization, discrete symmetry transformations and inner products} in free non-Hermitian scalar quantum field theories with $\mathcal{PT}$ symmetry, focusing on a prototype model of two complex scalar fields with anti-Hermitian mass mixing. Whereas the definition of the inner product is unique 
for theories described by Hermitian Hamiltonians, 
its formulation is not unique for non-Hermitian 
Hamiltonians. Energy eigenstates are not orthogonal with respect to the conventional Dirac inner product, so we must consider additional discrete transformations to define
a positive-definite norm.
We clarify the relationship between canonical-conjugate operators and introduce the additional discrete symmetry $\mathcal{C'}$, previously introduced for quantum-mechanical systems, and show that the $\mathcal{C'PT}$ 
inner product does yield a positive-definite norm,
and hence is appropriate for defining the Fock space in non-Hermitian models 
with $\mathcal{PT}$ symmetry in terms of energy eigenstates. We also discuss similarity transformations between $\mathcal{PT}$-symmetric
non-Hermitian scalar quantum field theories 
and Hermitian theories, showing that they
would require modification in the presence of interactions.
As an illustration of our discussion,
we compare particle mixing in a Hermitian theory and in the corresponding non-Hermitian model 
with $\mathcal{PT}$ symmetry, showing how the latter maintains unitarity and 
exhibits mixing between scalar and pseudoscalar bosons.}
~\\
\footnotesize{KCL-PH-TH/2020-19, CERN-TH-2020-063}
\\
\footnotesize{This is an author-prepared post-print of \href{https://doi.org/10.1103/PhysRevD.102.125030}{Phys.\ Rev.\ D {\bf 102} (2020) 125030}, published by the American Physical Society under the terms of the \href{https://creativecommons.org/licenses/by/4.0/}{CC BY 4.0} license (funded by SCOAP\textsuperscript{3}).}
\end{abstract}

\maketitle


\section{Introduction}

Recent years have witnessed growing interest in non-Hermitian quantum theories~\cite{AGU}, particularly those with $\mathcal{PT}$ symmetry,
where $\mathcal{P}$ and $\mathcal{T}$ denote parity and time reversal, respectively~\cite{Bender:2005tb}.
It is known that a quantum system described by a non-Hermitian Hamiltonian has real energies and leads to a unitary time evolution 
if this Hamiltonian and its eigenstates are invariant under $\mathcal{PT}$ symmetry \cite{Bender:1998ke}.
This increasing interest has been driven in part by theoretical analyses supporting the consistency of such theories in the context of both
quantum mechanics and quantum field theory, and in part by the realization that such theories have applications in many physical
contexts, e.g., photonics~\cite{Longhi, El-Ganainy} and phase transitions~\cite{Ashida, Matsumoto:2019are}. 
Although there are strong arguments for the consistency of $\mathcal{PT}$-symmetric quantum field
theory, a number of theoretical issues merit further attention. These include the analysis of discrete symmetries, which requires
in turn a careful analysis of the Fock spaces of non-Hermitian quantum field theories with $\mathcal{PT}$ symmetry
and their inner products.~\footnote{A detailed  description of the $\mathcal{PT}$ inner product in quantum mechanics can be found in Ref.~\cite{Mannheim:2017apd}.}

In this paper, we study and clarify these issues in the context of a minimal non-Hermitian bosonic 
field theory with $\mathcal{PT}$ symmetry at the classical and second-quantized levels. We construct explicitly
in the quantum version the operators generating discrete symmetries, and discuss the properties of candidate
inner products in Fock space. We also construct {a} similarity {transformation} between the free-field
$\mathcal{PT}$-symmetric non-Hermitian {model} and {the} corresponding Hermitian {counterpart}, showing explicitly
that the correspondence would not hold without modification in the presence of interactions.

As an application of this formalism, we discuss the simplest non-trivial 
prototype quantum particle system, namely 
mixing in models of non-interacting bosons --- building upon the study~\cite{AMS} that described how to interpret 
the corresponding $\mathcal{PT}$-symmetric Lagrangian.~\footnote{Self-interactions of these scalar fields
were considered in Ref.~\cite{AEMS1}, their coupling to an Abelian gauge field in Ref.~\cite{AEMS2} and to
non-Abelian gauge fields in Ref.~\cite{AEMS3}.
See Ref.~\cite{FTMono} for a study of 't Hooft-Polyakov monopoles in a non-Hermitian model.}
These systems appear in various physical situations of phenomenological
interest, such as coupled pairs of neutral mesons,
and also appear in the
$\mathcal{PT}$-symmetric extension of supersymmetry~\cite{AEM4}.
Issues arising in the formulation of such
theories include the roles of
discrete symmetries, the relationship  between the descriptions of mixing in the $\mathcal{PT}$-symmetric non-Hermitian 
case and the standard Hermitian case,~\footnote{See Refs.~\cite{Mannheim,FT1,FT2,Fring:2020bvr} 
for an alternative description of these models in terms of similarity transformations that map to a Hermitian model.}
and the status of unitarity, which has been questioned in non-Hermitian theories~\cite{Ohlsson:2019noy, OZmixing}.
As an example, we exhibit a mechanism allowing oscillations between scalar and pseudoscalar bosons,
which is possible with a mass-mixing matrix that is anti-Hermitian, but with real eigenvalues, and we compare with results in the previous literature.

The layout of our paper is as follows. In Sec.~\ref{sec:prototypes}, we introduce the minimal two-flavour non-Hermitian bosonic field theory that we study, discussing in Subsec.~\ref{sec:cdiscrete} its discrete symmetries $\mathcal{P}, \mathcal{T}$
and $\mathcal{C}'$~\cite{Bender:2002vv} at the classical level as well as the similarity transformation
relating it to a Hermitian theory, and mentioning a formal analogy with (1+1)-dimensional Special Relativity in Subsec.~\ref{sec:analogy}.
We discuss in Sec.~\ref{sec:quant} the second quantization of the theory in both the flavour and mass bases. Then, in Sec.~\ref{sec:discrete}, we discuss the
quantum versions of the discrete symmetries
and various definitions of the inner product in Fock space. In particular, we discuss
in Subsecs.~\ref{sec:parity} and \ref{sec:Cprime} the parity and $\mathcal{C}^\prime$
transformations, and we discuss the similarity transformation in Subsec.~\ref{sec:similarity}, emphasising that the equivalence between the {\it non-interacting} non-Hermitian model and a Hermitian theory does not in general
carry over to an {\it interacting} theory, in the absence of modifications. (Appendix~\ref{sec:alternativesim} compares the similarity transformation discussed in this paper
with a previous proposal~\cite{Mannheim} in the literature.)
In Subsec.~\ref{sec:products}, we distinguish the $\mathcal{PT}$ and $\mathcal{C'PT}$ inner products from the conventional Dirac inner product, showing that only the $\mathcal{C'PT}$ inner product is orthogonal and consistent with a positive-definite norm.\footnote{For an alternative approach, see Ref.~\cite{Mannheim:2017apd}.}
Subsection~\ref{sec:parity2} revisits the parity transformation, and, in Subsec.~\ref{sec:time}, we discuss time reversal in the light of our approach.
As an illustration, we discuss {in Sec.~\ref{sec:scalars} scalar-pseudoscalar mixing and oscillations in the non-Hermitian model, which reflect the fact that the parity operator does not commute with the Hamiltonian. We compare with oscillations in a Hermitian model and emphasize} that unitarity is respected. Our conclusions are summarized in Sec.~\ref{sec:conx}. 

A summary of notation is provided in Table~\ref{notation_table},
and some useful expressions are gathered in Appendix~\ref{sec:useful}.

\begin{table}
    \centering
    \begin{tabular}{c|c}
         $\ast$ & complex conjugation\\
         $\mathsf{T}$ & operator/matrix transposition \\
         $\mathcal{C}$ ($\hat{\mathcal{C}}$) & charge conjugation (operator)\\
         $\mathcal{C}'$ ($\hat{\mathcal{C}}$) & $\mathcal{C}'$ transformation (operator)\\
         $\mathcal{P}$ ($\hat{\mathcal{P}}$) & parity transformation (operator)\\
         $\mathcal{T}$ ($\hat{\mathcal{T}}$) & time-reversal transformation (operator)\\
         $\dag\equiv \ast\circ\mathsf{T}$ & Hermitian conjugation \\
         $\ddag\equiv \mathcal{PT}\circ \mathsf{T}$ & $\mathcal{PT}$ conjugation\\
         $\S\equiv \mathcal{C}'\mathcal{PT}\circ \mathsf{T}$ & $\mathcal{C}'\mathcal{PT}$ conjugation
    \end{tabular}
    \caption{Summary of notational conventions used in this article.}
    \label{notation_table}
\end{table}
 
 \section{Prototype Model}
 \label{sec:prototypes}
  \label{sec:bosonicmodel}
 
For definiteness, we frame the discussions that follow in the context of a prototype non-Hermitian but $\mathcal{PT}$-symmetric 
non-interacting bosonic field theory, comprising two flavours of complex spin-zero fields $\phi_i$ ($i=1,2$ are flavour indices) with non-Hermitian mass mixing. The two complex fields have four degrees of freedom, the minimal number needed to realize a non-Hermitian, 
$\mathcal{PT}$-symmetric field theory with real Lagrangian parameters. This should be contrasted with other non-Hermitian quantum field theories that have been discussed in the literature, which instead have fewer degrees of freedom but complex Lagrangian parameters~\cite{Blencowe:1997sy, Bender:2004vn, Bender:2004sa, Shalaby:2006fh, Bender:2012ea, Bender:2013qp, Shalaby:2009xda}.  It is understood that we are working in $3+1$-dimensional Minkowski spacetime
throughout.

The Lagrangian of the model is~\cite{AMS}
\begin{equation}
\label{eq:scalLag}
\mathcal{L}=\partial_{\nu}\phi_i^{*}\partial^{\nu}\phi_i-m_i^2\phi_i^{*}\phi_i-\mu^2(\phi_1^{*}\phi_2-\phi_2^{*}\phi_1)~,
\end{equation}
where $m_i^2>0$ and $\mu^2$ are real squared-mass parameters. The squared mass matrix
\begin{equation}
m^2\equiv \begin{pmatrix} m_1^2 & \mu^2 \\ -\mu^2 & m_2^2 \end{pmatrix}
\end{equation}
is skew symmetric. The squared eigenmasses are
\be
m^2_\pm=\frac{m_1^2+m_2^2}{2}\pm\frac{1}{2}\sqrt{(m_1^2-m_2^2)^2-4\mu^4}~,
\label{evalues}
\ee
which are real so long as
\be
\label{eq:etadef}
\eta\equiv\left|\frac{2\mu^2}{m_1^2-m_2^2}\right|\le1~,
\ee 
which defines the $\mathcal{PT}$-symmetric regime we consider here. 
For $\eta>1$, $\mathcal{PT}$ symmetry is broken by the complex eigenstates of the mass matrix;
the eigenmasses are not real and time evolution is not unitary. At $\eta=1$, the eigenvalues merge and the mass matrix becomes defective; at this \emph{exceptional point}, the squared mass matrix only has a single eigenvector (see, e.g., Ref.~\cite{AEMS3}). Hereafter, we take $m_1^2>m_2^2$, without loss of generality, so that we can omit the absolute value on the definition of the non-Hermitian parameter $\eta$ in Eq.~\eqref{eq:etadef}.

{By virtue of the non-Hermiticity of the Lagrangian, namely that $\mathcal{L}^{*}\neq \mathcal{L}$, the equations of motion obtained by varying the corresponding action with respect to $\phi^{\dag}\equiv(\phi_1^*,\phi_2^*)$ and $\phi\equiv(\phi_1,\phi_2)^{\mathsf{T}}$ differ by $\mu^2\to -\mu^2$, and therefore differ except for trivial solutions. However, we are free to choose either of these equations of motion to define the dynamics of the theory, since physical observables consistent with the $\mathcal{PT}$ symmetry of the model depend only on $\mu^4$~\cite{AMS}. As we show in this article, the choice of the equations of motion coincides with the choice of whether to take the Hamiltonian operator $\hat{H}(\mu^2)$ or $\hat{H}(-\mu^2)=\hat{H}^{\dag}(\mu^2)\neq \hat{H}(\mu^2)$ (its Hermitian conjugate) to generate the time evolution. For definiteness, and throughout this work, the classical dynamics of this theory will be defined by varying with respect to $\phi^{\dag}$, leading to the equations of motion
\begin{subequations}
 \begin{align}
     \Box \phi_i(x)+m_{ij}^2\phi_j(x)&=0~,\\
     \Box \phi_i^{*}(x)+m_{ij}^2\phi_j^{*}(x)&=0~.
 \end{align}
\end{subequations}
We reiterate that this choice amounts to no more than fixing the irrelevant overall sign of the mass-mixing term in Eq.~\eqref{eq:scalLag}.}

\subsection{Discrete Symmetries}
\label{sec:cdiscrete}

At the classical level with $c$-number Klein-Gordon fields, the Lagrangian in Eq.~\eqref{eq:scalLag} is 
$\mathcal{PT}$ symmetric under the {naive} transformations
\begin{subequations}
\label{eq:transfonaive}
\begin{align}
\mathcal{P}:\qquad &\phi_1(t,\mathbf{x})\to \phi_1'(t,-\mathbf{x})=+\phi_1(t,\mathbf{x})~,\nonumber\\
&\phi_2(t,\mathbf{x})\to \phi_2'(t,-\mathbf{x})=-\phi_2(t,\mathbf{x})~,\\
\mathcal{T}:\qquad &\phi_1(t,\mathbf{x})\to \phi_1'(-t,\mathbf{x})=+\phi_1^*(t,\mathbf{x})~,\nonumber\\
&\phi_2(t,\mathbf{x})\to \phi_2'(-t,\mathbf{x})=+\phi_2^*(t,\mathbf{x})~,
\end{align}
\end{subequations}
if one of the fields transforms as a scalar and the other as a pseudoscalar. 
As we show in this work, the Lagrangian of this model is also $\mathcal{PT}$ symmetric
at the quantum operator level.

However, it is important to realise that the Lagrangian in Eq.~\eqref{eq:scalLag}, and the resulting equations of motion, is not invariant under parity. In fact, the action of parity interchanges the two possible choices of equation of motion obtainable from Eq.~\eqref{eq:scalLag}. Taking this into account, there are a further two classical Lagrangians that are physically equivalent to Eq.~\eqref{eq:scalLag} and for which the parity transformation can be consistently defined:
\begin{subequations}
\label{eq:scalLagtilde}
\begin{align}
    \label{eq:scalLagtilde1}
    \tilde{\mathcal{L}}=\partial_{\nu}\tilde{\phi}_i^{*}\partial^{\nu}\phi_i-m_i^2\tilde{\phi}_i^{*}\phi_i-\mu^2(\tilde{\phi}_1^{*}\phi_2-\tilde{\phi}_2^{*}\phi_1)~,\\
    \tilde{\mathcal{L}}^*=\partial_{\nu}\phi_i^{*}\partial^{\nu}\tilde{\phi}_i-m_i^2\phi_i^{*}\tilde{\phi}_i+\mu^2(\phi_1^{*}\tilde{\phi}_2-\phi_2^{*}\tilde{\phi}_1)~,
\end{align}
\end{subequations}
and their ``tilde'' conjugates
\begin{subequations}
\label{eq:scalLagtildetilde}
\begin{align}
    \tilde{\tilde{\mathcal{L}}}=\partial_{\nu}\phi_i^{*}\partial^{\nu}\tilde{\phi}_i-m_i^2\phi_i^{*}\tilde{\phi}_i-\mu^2(\phi_1^{*}\tilde{\phi}_2-\phi_2^{*}\tilde{\phi}_1)~,\\
    \tilde{\tilde{\mathcal{L}}}^*=\partial_{\nu}\tilde{\phi}_i^{*}\partial^{\nu}\phi_i-m_i^2\tilde{\phi}_i^{*}\phi_i+\mu^2(\tilde{\phi}_1^{*}\phi_2-\tilde{\phi}_2^{*}\phi_1)~,
\end{align}
\end{subequations}
differing by $\mu^2\to-\mu^2$, i.e., $\tilde{\mathcal{L}}(\mu^2)=\tilde{\tilde{\mathcal{L}}}^*(-\mu^2)$. The fields indicated by a tilde are defined by the action of parity, namely
\begin{align}
\label{eq:revisedP}
\mathcal{P}:&\quad \phi_1(t,x)\to \phi_1'(t,-\mathbf{x})=+\tilde{\phi}_1(t,\mathbf{x})~,\nonumber\\
&\quad \phi_2(t,x)\to \phi_2'(t,-\mathbf{x})=-\tilde{\phi}_2(t,\mathbf{x})~.
\end{align}
For these Lagrangians, the Euler-Lagrange equations are self-consistent, and Eq.~\eqref{eq:scalLagtilde1} yields
\begin{subequations}
 \begin{align}
     \Box \tilde{\phi}_i(x)+m_{ji}^2\tilde{\phi}_j(x)&=0~,\\
     \Box \tilde{\phi}_i^{*}(x)+m_{ji}^2\tilde{\phi}_j^{*}(x)&=0~.
 \end{align}
\end{subequations}
Making use of Eq.~\eqref{eq:revisedP} and the time-reversal transformations in Eq.~\eqref{eq:transfonaive}, we see that the Lagrangians in Eqs.~\eqref{eq:scalLagtilde} and~\eqref{eq:scalLagtildetilde} remain $\mathcal{PT}$ symmetric.

In order to illustrate the flavour structure of this model, it is convenient to consider a matrix model with non-Hermitian squared Hamiltonian given by
\begin{equation}
    \label{eq:matmod}
    H^2=\begin{pmatrix} m_1^2 & \mu^2 \\ -\mu^2 & m_2^2\end{pmatrix}~,
\end{equation}
reflecting the squared mass matrix of the model in Eq.~\eqref{eq:scalLag}. The Hamiltonian is (up to an overall sign)
\begin{equation}
    \label{eq:H}
    H=\frac{1}{\sqrt{m_1^2+m_2^2{\  \pm \ }2\sqrt{m_1^2m_2^2+\mu^4}}}\begin{pmatrix} m_1^2\pm\sqrt{m_1^2m_2^2+\mu^4} & \mu^2 \\ -\mu^2 & m_2^2\pm\sqrt{m_1^2m_2^2+\mu^4}\end{pmatrix}~,
\end{equation}
with eigenvectors~\cite{AMS}
\be\label{masseigenstates}
{\bf e_+}=N\begin{pmatrix} \eta \\ -1+\sqrt{1-\eta^2} \end{pmatrix}~~~~,~~~~{\bf e_-}=N\begin{pmatrix} -1+\sqrt{1-\eta^2} \\ \eta \end{pmatrix}~,
\ee
where $N$ is a normalization factor. We remark that it is necessary to take the positive square root in Eq.~\eqref{eq:H} in order for the Hamiltonian to be well defined at the exceptional points.

{Under a parity transformation, the squared Hamiltonian transforms as
\begin{equation}
    \mathcal{P}:\quad PH^2P=\begin{pmatrix} m_1^2 & -\mu^2 \\ \mu^2 & m_2^2\end{pmatrix}=H^{2,\mathsf{T}}~,
\end{equation}
where the matrix $P$ is a $2\times2$ matrix that reflects the intrinsic parities of the scalar and pseudoscalar fields in Eq.~\eqref{eq:scalLag}:
\begin{equation}
P=\begin{pmatrix} 1 & 0 \\ 0 & -1\end{pmatrix}~.
\end{equation}}

An important difference from the Hermitian case is that
the eigenvectors \eqref{masseigenstates} are {\it not} orthogonal with respect to the Hermitian inner product, ${\bf e}_-^*\cdot{\bf e_+}\ne0$.
Instead, they are orthogonal with respect to the $\mathcal{PT}$ inner product:
\begin{subequations}
\label{inner}
\begin{align}
\mathbf{e}_+^{\ddag}\mathbf{e}_+={\bf e}_+^\mathcal{PT}\cdot {\bf e}_+=-\mathbf{e}_-^{\ddag}\mathbf{e}_-=-{\bf e}_-^\mathcal{PT}\cdot {\bf e}_-=1~,\\
\mathbf{e}_+^{\ddag}\mathbf{e}_-={\bf e}_+^\mathcal{PT}\cdot {\bf e}_-=\mathbf{e}_-^{\ddag}\mathbf{e}_+={\bf e}_-^\mathcal{PT}\cdot {\bf e}_+=0~,
\end{align}
\end{subequations}
where $\ddag\equiv \mathcal{PT}\circ \mathsf{T}$, with $\mathsf{T}$ indicating matrix transposition,~\footnote{The $\ddag$ notation was introduced in Ref.~\cite{Bender:1998gh} and extended in Ref.~\cite{AMS}.} and 
\be
\label{eq:paritymatrix}
{\bf e}_{\pm}^\mathcal{PT}=P{\bf e}_{\pm}^\ast~,
\ee
and we choose the normalization constant~\cite{AMS}
\be
N=(2\eta^2-2+2\sqrt{1-\eta^2})^{-1/2}~.
\ee
Notice, however, that one of the eigenvectors, viz.~$\mathbf{e}_-$, has negative $\mathcal{PT}$ norm, as is expected for a non-Hermitian $\mathcal{PT}$-symmetric theory. Note that the Hamiltonian is $\mathcal{PT}$ symmetric in the sense that $[H,\ddag]=0$.

As was first shown in Ref.~\cite{Bender:2002vv}, the $\mathcal{PT}$ symmetry of the Hamiltonian allows the construction of an additional symmetry transformation, which we denote by $\mathcal{C}'$ and which can be used to construct a positive-definite norm:~the $\mathcal{C}'\mathcal{PT}$ norm.~\footnote{As we discuss in Subsec.~\ref{sec:Cprime}, the $\mathcal{C}'$ transformation in a $\mathcal{PT}$-symmetric quantum field theory
cannot be identified with charge conjugation.}

The $C'$ matrix for the squared Hamiltonian in Eq.~\eqref{eq:matmod} is given by~\cite{AEMS1}
\begin{equation}
\label{eq:Cprimescalmat}
C'=RPR^{-1} \equiv \frac{1}{\sqrt{1-\eta^2}}\begin{pmatrix}1 & -\eta \\ \eta & -1 \end{pmatrix}~,
\end{equation}
where
\begin{equation}
    R \equiv N\begin{pmatrix} \eta & 1-\sqrt{1-\eta^2} \\ 1-\sqrt{1-\eta^2} & \eta\end{pmatrix}
\end{equation}
gives the matrix similarity transformation that diagonalizes the Hamiltonian, i.e.,
\begin{equation}
    \label{eq:Rtrans}
    h^2=R H^2R^{-1}=\begin{pmatrix} m_+^2 & 0 \\ 0 & m_-^2\end{pmatrix}~.
\end{equation}
We note that this similarity transformation leads to a \emph{Hermitian} Hamiltonian. Indeed, it is well established that for \emph{non-interacting} non-Hermitian $\mathcal{PT}$-symmetric theories the $\mathcal{C}'$ transformation is directly related to the similarity transformation that maps the theory to a Hermitian one. Specifically, the matrix $C'$ can be written in the form~\footnote{There is a relative sign in the definition of the matrix $Q$ compared with Refs.~\cite{Bender:2004vn, Bender:2004sa}, due to differing conventions for the definition of the $\mathcal{C}'\mathcal{PT}$ inner product.}
\begin{equation}
    \label{eq:matrixCprimeidentity}
    C'=e^{-Q}P~,
\end{equation}
where the matrix $Q$ has the property that
\begin{equation}
    \label{eq:similarity}
    h^2=e^{-Q/2}H^2e^{Q/2}~,
\end{equation}
leading to the same Hermitian Hamiltonian. Using the identity
\begin{equation}
R=PR^{-1}P~,    
\end{equation}
we can confirm that Eq.~\eqref{eq:similarity} is consistent with Eq.~\eqref{eq:Rtrans}, i.e.,
\begin{equation}
    \label{eq:RQrel}
    e^{-Q}=C'P=RPR^{-1}P=R^2=\frac{1}{\sqrt{1-\eta^2}}\begin{pmatrix} 1 & \eta \\ \eta & 1\end{pmatrix}~,
\end{equation}
and it follows that
\begin{equation}
    Q=\ln R^{-2}=-{\rm arctanh}\left(\eta\right)\bar{Q}~,
\end{equation}
where
\begin{equation}
\label{eq:barQdef}
\bar{Q}\equiv \begin{pmatrix}0 & 1 \\ 1 & 0 \end{pmatrix}~.
\end{equation}
The $\mathcal{C}'\mathcal{PT}$ conjugates of the eigenvectors are
\begin{subequations}
\begin{align}
\mathbf{e}_{+}^{\mathcal{C}'\mathcal{PT}}&=C'P\mathbf{e}_{+}=\frac{1}{\sqrt{1-\eta^2}}\begin{pmatrix} 1 & \eta \\ \eta & 1\end{pmatrix}\mathbf{e}_+=N\begin{pmatrix} \eta \\ 1-\sqrt{1-\eta^2}\end{pmatrix}~,\\
\mathbf{e}_{-}^{\mathcal{C}'\mathcal{PT}}&=C'P\mathbf{e}_{-}=\frac{1}{\sqrt{1-\eta^2}}\begin{pmatrix} 1 & \eta \\ \eta & 1\end{pmatrix}\mathbf{e}_-=N\begin{pmatrix} 1-\sqrt{1-\eta^2} \\ \eta\end{pmatrix}~,
\end{align}
\end{subequations}
and it is easy to check that their $\mathcal{C}'\mathcal{PT}$ norms are positive definite:
\begin{equation}
    \mathbf{e}_{\pm}^{\S}\mathbf{e}_{\pm}=\mathbf{e}_{\pm}^{\mathcal{C}'\mathcal{PT}}\cdot \mathbf{e}_{\pm}=1~,
\end{equation}
where $\S\equiv \mathcal{C}'\mathcal{PT}\circ \mathsf{T}$, and that they are orthogonal:
\begin{equation}
    \mathbf{e}_{\pm}^{\mathcal{C}'\mathcal{PT}}\cdot \mathbf{e}_{\mp}=0~.
\end{equation}
We note that $C'$ reduces to $P$ in the Hermitian limit $\eta\to 0$, so that the $\mathcal{C}'\mathcal{PT}$ inner product reduces to the Hermitian inner product.

It will prove helpful to note that we can also write the mass eigenstates and their $\mathcal{C}'\mathcal{PT}$ conjugates in the following ways:
\begin{subequations}
\label{eq:CprimePversusR}
 \begin{align}
    \mathbf{e}_+=R^{-1}\mathbf{e}_1=R^{-1}_{1j}\mathbf{e}_j~,\\
    \mathbf{e}_-=R^{-1}\mathbf{e}_2=R^{-1}_{2j}\mathbf{e}_j~,\\
     C'P\mathbf{e}_+=R\mathbf{e}_1=R_{1j}\mathbf{e}_j~,\\
     C'P\mathbf{e}_-=R\mathbf{e}_2=R_{2j}\mathbf{e}_j~,
 \end{align}
\end{subequations}
where
\begin{equation}
    \mathbf{e}_1=\begin{pmatrix} 1 \\ 0 \end{pmatrix} \qquad \text{and}\qquad \mathbf{e}_2=\begin{pmatrix} 0 \\ 1 \end{pmatrix}
\end{equation}
are the flavour eigenstates. In addition, we can show that
\begin{subequations}
\begin{align}
    \mathbf{e}_+^{\mathsf{T}}C'P\mathbf{e}_+=\mathbf{e}_1^{\mathsf{T}}R^{-1}C'PR^{-1}\mathbf{e}_{1}=\mathbf{e}_1\cdot\mathbf{e}_1~,\\
    \mathbf{e}_-^{\mathsf{T}}C'P\mathbf{e}_-=\mathbf{e}_2^{\mathsf{T}}R^{-1}C'PR^{-1}\mathbf{e}_{2}=\mathbf{e}_2\cdot\mathbf{e}_2~,
\end{align}
\end{subequations}
i.e., the Hermitian inner product of the flavour eigenstates, which is not problematic, is related to the $\mathcal{C}'\mathcal{PT}$ inner product of the mass eigenstates.

\subsection{Analogy with 1+1-Dimensional Special Relativity}
\label{sec:analogy}

The similarity transformation \eqref{eq:similarity} between the flavour and mass eigenbases is not a rotation, since the original mass-mixing matrix 
is not Hermitian. Interestingly, however, it is analogous to a Lorentz boost in the 1+1-dimensional field space $(\phi_1,\phi_2)$ with metric $P$. 
Indeed, one can easily check that $R$ can be written in the form
\be
R=\gamma\begin{pmatrix} 1 & v \\ v & 1 \end{pmatrix}~,
\ee
where 
\be
v\equiv\frac{1}{\eta}\left(1-\sqrt{1-\eta^2}\right)\qquad \text{and}\qquad \gamma=\frac{1}{\sqrt{1-v^2}}~.
\ee
The $\mathcal{PT}$-symmetric phase, characterized by $0\le\eta\le1$, corresponds to the ``subluminal regime'' $0\le v\le1$, whereas the 
$\mathcal{PT}$ symmetry-breaking phase corresponds to the ``superluminal regime'' $v>1$.

As is known from Special Relativity, the Pauli matrix $\sigma_1$ generates 1+1-dimensional Lorentz boosts, and one can also write
\be
R=\exp(\alpha\sigma_1)\qquad \text{with}\qquad \alpha\equiv{\rm arctanh}\,v~,
\ee
which is consistent with Eqs.~\eqref{eq:RQrel} to~\eqref{eq:barQdef}, since $\bar{Q}=\sigma_1$ and
\begin{equation}
    {\rm arctanh}\,v=\frac{1}{2}\,{\rm arctanh}\,\eta~.
\end{equation}
The quadratic field invariants under a change of basis are 
$\phi^{\dag}_i P_{ij}\phi_j$ and $\phi_i P_{ij}\phi_j$, as well as their complex conjugates.

\section{Quantization}
\label{sec:quant}
Having understood the flavour structure of this non-Hermitian model, we now turn our attention to its second quantization.

\subsection{Flavour Basis}
\label{flavour}

For the two-flavour model, the mass matrix is not diagonal in the flavour basis, and the same is true of the energy, whose square is given by
\begin{equation}
    E_{ij}^2(\mathbf{p})=\mathbf{p}^2\delta_{ij}+m^2_{ij}~.
\end{equation}
Since the squared mass matrix $m^2$ is non-Hermitian, so too is the energy, i.e., $E^{\dag}\neq E$.

As described earlier, and due to the non-Hermiticity of the action, we obtain distinct but physically equivalent equations of motion by varying with respect to $\hat{\phi}_i^{\dag}$ or $\hat{\phi}_i$ (see, e.g., Ref.~\cite{AMS}). Starting from the Lagrangian
\begin{equation}
    \label{eq:scalLaghat}
    \hat{\mathcal{L}}=\partial_{\nu}\hat{\phi}_i^{\dag}\partial^{\nu}\hat{\phi}_i-m_i^2\hat{\phi}_i^{\dag}\hat{\phi}_i-\mu^2\left(\hat{\phi}_1^{\dag}\hat{\phi}_2-\hat{\phi}_2^{\dag}\hat{\phi}_1\right)~,
\end{equation}
and choosing the equations of motion by varying with respect to $\hat{\phi}_i^{\dag}$, we have
\begin{subequations}
\label{eq:eoms}
 \begin{align}
     \Box \hat{\phi}_i+m^2_{ij}\hat{\phi}_j&=0~,\\
     \Box \hat{\phi}^{\dag}_i+m^2_{ij}\hat{\phi}^{\dag}_j&=0~.
 \end{align}
\end{subequations}
Since $E^{\dag}_{ij}=E_{ji}$, it follows that the plane-wave decompositions of the scalar field operators are
\begin{subequations}
\label{eq:scalarfielddefs}
\begin{align}
     \hat{\phi}_i(x)&=\int_{\mathbf{p}}\left[2E(\mathbf{p})\right]^{-1/2}_{ij}\left[\left(e^{-ip\cdot x}\right)_{jk}\hat{a}_{k,\mathbf{p}}(0)+\left(e^{ip\cdot x}\right)_{jk}\hat{c}^{\dag}_{k,\mathbf{p}}(0)\right]~,\\
     \hat{\phi}_i^{\dag}(x)&=\int_{\mathbf{p}}\left[2E(\mathbf{p})\right]^{-1/2}_{ij}\left[\left(e^{-ip\cdot x}\right)_{jk}\hat{c}_{k,\mathbf{p}}(0)+\left(e^{ip\cdot x}\right)_{jk}\hat{a}^{\dag}_{k,\mathbf{p}}(0)\right]~,
\end{align}
\end{subequations}
where we have used the shorthand notation
\begin{equation}
    \int_{\mathbf{p}}\equiv \int\frac{\mathrm{d}^3 \mathbf{p}}{(2\pi)^3}
\end{equation}
for the three-momentum integral. Since the energy is a rank-two tensor in flavour space, it follows that the energy factor in the phase-space measure and the plane-wave factors must also be rank-two tensors in flavour space, with the matrix-valued exponentials being understood in terms of their series expansions.~\footnote{For a comprehensive discussion of flavour covariance, see Ref.~\cite{Dev:2014laa}. For notational simplicity,
we do not distinguish in the present work between covariant and contravariant indices in flavour space.} 

{We have normalised the particle and antiparticle creation operators $\hat{a}^{\dag}$ and $\hat{c}^{\dag}$, and the annihilation operators $\hat{a}$ and $\hat{c}$, such that they have mass dimension $-3/2$. As a result, their canonical commutation relations (with respect to Hermitian conjugation) are isotropic both in the flavour and mass eigenbases at the initial time surface for the quantization, viz.~$t=0$. Specifically, we have
\begin{equation}
    \label{eq:ladderCCR}
    \left[\hat{a}_{i,\mathbf{p}}(0),\hat{a}^{\dag}_{j,\mathbf{p}'}(0)\right]=\left[\hat{c}_{i,\mathbf{p}}(0),\hat{c}^{\dag}_{j,\mathbf{p}'}(0)\right]=(2\pi)^3\delta_{ij}\delta^{3}(\mathbf{p}-\mathbf{p}')~.
\end{equation}
However, the non-orthogonality of the Hermitian inner product becomes manifest at different times:
\begin{align}
    \label{eq:ladderCCRt}
    &\left[\hat{a}_{i,\mathbf{p}}(t),\hat{a}^{\dag}_{j,\mathbf{p}'}(t)\right]=\left[\hat{c}_{i,\mathbf{p}}(t),\hat{c}^{\dag}_{j,\mathbf{p}'}(t)\right]=(2\pi)^3\big(e^{-iE_{\mathbf{p}}t}\big)_{ik}\big(e^{iE^{\mathsf{T}}_{\mathbf{p}'}t}\big)_{kj}\delta^{3}(\mathbf{p}-\mathbf{p}')\nonumber\\&=(2\pi)^3\delta^{3}(\mathbf{p}-\mathbf{p}')\begin{cases} 1+\frac{4\mu^4}{(m_1^2-m_2^2)^2-4\mu^4}\left[1-\cos\left(\frac{\sqrt{(m_1^2-m_2^2)^2-4\mu^4}t}{\sqrt{2}\bar{E}_{\mathbf{p}}}\right)\right]~,& i=j\\ -\frac{2\mu^2(m_1^2-m_2^2)}{(m_1^2-m_2^2)^2-4\mu^4}\left[1-\cos\left(\frac{\sqrt{(m_1^2-m_2^2)^2-4\mu^4}t}{\sqrt{2}\bar{E}_{\mathbf{p}}}\right)\right.\\ \ \left.+(-) i\sqrt{1-\frac{4\mu^4}{(m_1^2-m_2^2)^2}}\sin\left(\frac{\sqrt{(m_1^2-m_2^2)^2-4\mu^4}t}{\sqrt{2}\bar{E}_{\mathbf{p}}}\right)\right]~,& i=1(2)~,~j=(2)1\end{cases},
\end{align}
where
\begin{equation}
    \bar{E}_{\mathbf{p}}=\left[\mathbf{p}^2+\frac{m_1^2+m_2^2}{2}+\sqrt{\left(\mathbf{p}^2+m_1^2\right)\left(\mathbf{p}^2+m_2^2\right)+\mu^4}\right]^{1/2}~,
\end{equation}
and it is clear that the canonical-conjugate variables cannot be related by Hermitian conjugation.
}

{As identified earlier,} the non-Hermitian terms of the Lagrangian in Eq.~\eqref{eq:scalLag} violate parity. In fact, parity acts to transform the Lagrangian in Eq.~\eqref{eq:scalLag} and the corresponding Hamiltonian into their Hermitian conjugates. As a result, the field operators and their parity conjugates evolve with respect to $\hat{H}$ and $\hat{H}^{\dag}$ respectively. To account for this, it is convenient to introduce a second pair of field operators, denoted by a check ($\check{\ }$), which satisfy the alternative choice of equations of motion:
 \begin{subequations}
 \label{eq:eomsPconjugate}
  \begin{align}
      \Box \check{\phi}_i(x)+(m^2)^{\mathsf{T}}_{ij}\check{\phi}_j(x)&=0~,\\
      \Box \check{\phi}^{\dag}_i(x)+(m^2)^{\mathsf{T}}_{ij}\check{\phi}^{\dag}_j(x)&=0~,
  \end{align}
 \end{subequations}
and are related to $\hat{\phi}_i(x)$ and $\hat{\phi}^{\dag}_i(x)$ by parity:
 \begin{subequations}
 \label{eq:phivarrels}
  \begin{align}
      P_{ij}\check{\phi}_j(\mathcal{P}x)&=\hat{\mathcal{P}}\hat{\phi}_i(x)\hat{\mathcal{P}}^{-1}~,\\
      P_{ij}\check{\phi}_j^{\dag}(\mathcal{P}x)&=\hat{\mathcal{P}}\hat{\phi}_i^{\dag}(x)\hat{\mathcal{P}}^{-1}~,
  \end{align}
 \end{subequations}
cf.~Eq.~\eqref{eq:revisedP}. Their plane-wave decompositions are
\begin{subequations}
\label{eq:scalarfieldvardefs}
\begin{align}
     \check{\phi}_i(x)&=\int_{\mathbf{p}}\left[2E^{\mathsf{T}}(\mathbf{p})\right]^{-1/2}_{ij}\left[\left(e^{-ip^{\mathsf{T}}\cdot x}\right)_{jk}{\check{a}_{k,\mathbf{p}}}(0)+\left(e^{ip^{\mathsf{T}}\cdot x}\right)_{jk}{\check{c}^{\dag}_{k,\mathbf{p}}}(0)\right]~,\\
     \check{\phi}_i^{\dag}(x)&=\int_{\mathbf{p}}\left[2E^{\mathsf{T}}(\mathbf{p})\right]^{-1/2}_{ij}\left[\left(e^{-ip^{\mathsf{T}}\cdot x}\right)_{jk}{\check{c}_{k,\mathbf{p}}}(0)+\left(e^{ip^{\mathsf{T}}\cdot x}\right)_{jk}{\check{a}^{\dag}_{k,\mathbf{p}}}(0)\right]~,
\end{align}
\end{subequations}
where $[p^{\mathsf{T}}\cdot x]_{ij}=E^{\mathsf{T}}_{ij}\cdot x^0-\delta_{ij}\mathbf{p}\cdot\mathbf{x}$, differing from Eq.~\eqref{eq:scalarfielddefs} by $E\to E^{\mathsf{T}}$. We emphasize that $\hat{\phi}_i$ and $\check{\phi}_i^{\dag}$ evolve with $\hat{H}$, whereas $\hat{\phi}^{\dag}_i$ and $\check{\phi}_i$ evolve with $\hat{H}^{\dag}$. The relations between the creation and annihilation operators are analogous to Eq.~\eqref{eq:phivarrels}:
 \begin{subequations}
 \label{eq:acvarrels}
  \begin{align}
      P_{ij}\check{a}_{j,-\mathbf{p}}(t)&=\hat{\mathcal{P}}\hat{a}_{i,\mathbf{p}}(t)\hat{\mathcal{P}}^{-1}~,\\
      P_{ij}\check{a}_{j,-\mathbf{p}}^{\dag}(t)&=\hat{\mathcal{P}}\hat{a}_{i,\mathbf{p}}^{\dag}(t)\hat{\mathcal{P}}^{-1}~,
  \end{align}
 \end{subequations}
 and likewise for $\hat{c}_i$ and $\hat{c}_i^{\dag}$. We emphasise, however, that the distinction between checked and hatted operators is necessary only away from the initial time surface of the quantization; namely, we have
 \begin{equation}
    \check{a}^{(\dag)}_{i,\mathbf{p}}(0)=\hat{a}^{(\dag)}_{i,\mathbf{p}}(0)~,
\end{equation}
and likewise for the antiparticle operators. Making use of this identification, it is more illustrative to write the various field operators in the following forms:
\begin{subequations}
\label{eq:scalarfielddefscomplete}
\begin{align}
     \hat{\phi}_i(x)&=\int_{\mathbf{p}}\left[2E(\mathbf{p})\right]^{-1/2}_{ij}\left[\left(e^{-ip\cdot x}\right)_{jk}\hat{a}_{k,\mathbf{p}}(0)+\left(e^{ip\cdot x}\right)_{jk}\check{c}^{\dag}_{k,\mathbf{p}}(0)\right]~,\\
     \hat{\phi}_i^{\dag}(x)&=\int_{\mathbf{p}}\left[2E(\mathbf{p})\right]^{-1/2}_{ij}\left[\left(e^{-ip\cdot x}\right)_{jk}\check{c}_{k,\mathbf{p}}(0)+\left(e^{ip\cdot x}\right)_{jk}\hat{a}^{\dag}_{k,\mathbf{p}}(0)\right]~,\\
    \check{\phi}_i(x)&=\int_{\mathbf{p}}\left[2E^{\mathsf{T}}(\mathbf{p})\right]^{-1/2}_{ij}\left[\left(e^{-ip^{\mathsf{T}}\cdot x}\right)_{jk}{\check{a}_{k,\mathbf{p}}}(0)+\left(e^{ip^{\mathsf{T}}\cdot x}\right)_{jk}\hat{c}^{\dag}_{k,\mathbf{p}}(0)\right]~,\\
     \check{\phi}_i^{\dag}(x)&=\int_{\mathbf{p}}\left[2E^{\mathsf{T}}(\mathbf{p})\right]^{-1/2}_{ij}\left[\left(e^{-ip^{\mathsf{T}}\cdot x}\right)_{jk}\hat{c}_{k,\mathbf{p}}(0)+\left(e^{ip^{\mathsf{T}}\cdot x}\right)_{jk}{\check{a}^{\dag}_{k,\mathbf{p}}}(0)\right]~,
\end{align}
\end{subequations}
where we draw attention to the fact that particle and antiparticle operators appear with opposing hats and checks. This convention makes manifest the necessity for both the particle annihilation operator $\hat{a}$ and the antiparticle creation operator $\check{c}^{\dag}$, which appear in the field operator $\hat{\phi}$, to evolve with the Hamiltonian $\hat{H}$, and not with $\hat{H}$ and $\hat{H}^{\dag}$, respectively, as one might have expected naively.

{A canonical-conjugate pair of variables, e.g., $\hat{\phi}_i$ and $\hat{\pi}_i$, must evolve subject to the  same Hamiltonian, i.e., they must both evolve according to $\hat{H}$ or both according to $\hat{H}^{\dag}$. The conjugate momentum operators are therefore
\begin{subequations}
\label{eq:conjugatemoms}
 \begin{align}
     \hat{\pi}_i(x)&=\partial_t\check{\phi}_i^{\dag}(x)=-i\int_{\mathbf{p}}\left[2E^{\mathsf{T}}(\mathbf{p})\right]^{1/2}_{ij}\left[\left(e^{-ip^{\mathsf{T}}\cdot x}\right)_{jk}\hat{c}_{k,\mathbf{p}}(0)-\left(e^{ip^{\mathsf{T}}\cdot x}\right)_{jk}{\check{a}^{\dag}_{k,\mathbf{p}}}(0)\right]~,\\
     \hat{\pi}_i^{\dag}(x)&=\partial_t\check{\phi}_i(x)=-i\int_{\mathbf{p}}\left[2E^{\mathsf{T}}(\mathbf{p})\right]^{1/2}_{ij}\left[\left(e^{-ip^{\mathsf{T}}\cdot x}\right)_{jk}{\check{a}_{k,\mathbf{p}}}(0)-\left(e^{ip^{\mathsf{T}}\cdot x}\right)_{jk}\hat{c}^{\dag}_{k,\mathbf{p}}(0)\right]~.
 \end{align}
\end{subequations}
Were we instead to insist on the usual relationship between the conjugate momentum operator and the time derivative of the field operator, i.e., $\hat{\pi}_i=\partial_t\hat{\phi}_i^{\dag}$, we would force $\hat{\phi}_i$ and $\hat{\pi}_i$ to evolve with respect to $\hat{H}$ and $\hat{H}^{\dag}$, respectively, and they would therefore not be canonical-conjugate variables. We recover the usual relationship between the field and conjugate momentum only in the Hermitian limit $\mu\to 0$.} It may readily be confirmed that Eqs.~\eqref{eq:ladderCCR}, \eqref{eq:scalarfielddefscomplete} and~\eqref{eq:conjugatemoms} lead to canonical equal-time commutation relations
\begin{subequations}
\label{eq:hatCCR}
\begin{align}
    \big[\hat{\phi}_i(t,\mathbf{x}),\hat{\phi}^{\dag}_j(t,\mathbf{y})\big]&=0~,\\
    \big[\hat{\phi}_i(t,\mathbf{x}),\hat{\pi}_j(t,\mathbf{y})\big]& =  i\delta_{ij}\delta^{3}(\mathbf{x}-\mathbf{y})~,\\
    \big[\hat{\phi}^{\dag}_i(t,\mathbf{x}),\hat{\pi}^{\dag}_j(t,\mathbf{y})\big]& =  i\delta_{ij}\delta^{3}(\mathbf{x}-\mathbf{y})~,
\end{align}
\end{subequations}
In addition, we have that
\begin{subequations}
\label{eq:checkCCR}
\begin{align}
    \big[\hat{\phi}_i(t,\mathbf{x}),\check{\phi}^{\dag}_j(t,\mathbf{y})\big]&=0~,\\
    \big[\check{\phi}_i(t,\mathbf{x}),\check{\phi}^{\dag}_j(t,\mathbf{y})\big]&=0~,\\
    \big[\check{\phi}_i(t,\mathbf{x}),\check{\pi}_j(t,\mathbf{y})\big]& =  i\delta_{ij}\delta^{3}(\mathbf{x}-\mathbf{y})~,\\
    \big[\check{\phi}^{\dag}_i(t,\mathbf{x}),\check{\pi}^{\dag}_j(t,\mathbf{y})\big]& =  i\delta_{ij}\delta^{3}(\mathbf{x}-\mathbf{y})~,
\end{align}
\end{subequations}
where
\begin{subequations}
 \begin{align}
     \check{\pi}_i(x)&=\partial_t\hat{\phi}_i^{\dag}(x)=-i\int_{\mathbf{p}}\left[2E(\mathbf{p})\right]^{1/2}_{ij}\left[\left(e^{-ip\cdot x}\right)_{jk}\check{c}_{k,\mathbf{p}}(0)-\left(e^{ip\cdot x}\right)_{jk}{\hat{a}^{\dag}_{k,\mathbf{p}}}(0)\right]~,\\
     \check{\pi}_i^{\dag}(x)&=\partial_t\hat{\phi}_i(x)=-i\int_{\mathbf{p}}\left[2E(\mathbf{p})\right]^{1/2}_{ij}\left[\left(e^{-ip\cdot x}\right)_{jk}{\hat{a}_{k,\mathbf{p}}}(0)-\left(e^{ip\cdot x}\right)_{jk}\check{c}^{\dag}_{k,\mathbf{p}}(0)\right]~.
 \end{align}
\end{subequations}

We can now write down the Hamiltonian (density) operator that generates the time evolution consistent with the equations of motion in Eqs.~\eqref{eq:eoms} and \eqref{eq:eomsPconjugate}:
\begin{equation}
    \label{eq:hatHamiltonian}
    \hat{\mathcal{H}}=\check{\pi}^{\dag}_i(x)\hat{\pi}_i(x)+\bm{\nabla}\check{\phi}^{\dag}_i(x)\cdot \bm{\nabla}\hat{\phi}_i(x)+\check{\phi}^{\dag}_i(x)m_{ij}^2\hat{\phi}_j(x)~.
\end{equation}
The corresponding Lagrangian density is
\begin{equation}
    \hat{\mathcal{L}}=\partial_{\nu}\check{\phi}^{\dag}_i(x)\partial^{\nu}\hat{\phi}_i(x)-\check{\phi}^{\dag}_i(x)m_{ij}^2\hat{\phi}_j(x)~.
\end{equation}
Had we made the alternative choice for the equations of motion, i.e., varying the Lagrangian in Eq.~\eqref{eq:scalLaghat} with respect to $\hat{\phi}_i$, the time evolution would instead be generated by
\begin{equation}
    \label{eq:checkHamiltonian}
    \hat{\mathcal{H}}^{\dag}=\hat{\pi}^{\dag}_i(x)\check{\pi}_i(x)+\bm{\nabla}\hat{\phi}^{\dag}_i(x)\cdot \bm{\nabla}\check{\phi}_i(x)+\hat{\phi}^{\dag}_i(x)m_{ji}^2\check{\phi}_j(x)~,
\end{equation}
but the physical results would be identical.

\subsection{Mass Basis}
\label{sec:mass}

The transformation to the mass eigenbasis is effected by the similarity transformation
\begin{subequations}
\label{eq:Rtrans2}
 \begin{align}
     \hat{\xi}_i(x)&=R_{ij}\hat{\phi}_j(x)~,\\
     \check{\xi}^{\S}_i(x)&=\check{\phi}^{\dag}_j(x)R_{ji}^{-1}~.
 \end{align}
\end{subequations}
By virtue of Eq.~\eqref{eq:CprimePversusR}, or making use of the transformations defined in the next section, we can readily convince ourselves that the variables $\hat{\xi}_i$ and $\check{\xi}^{\S}_i$ are the $\mathcal{C}'\mathcal{PT}$-conjugate variables of the mass eigenbasis.

We infer from Eq.~\eqref{eq:Rtrans2} that particle annihilation and anti-particle creation operators have to transform in the same way, under both the similarity transformation to the mass eigenbasis and  $\mathcal{C}'$ (see Subsec.~\ref{sec:cdiscrete}).

\section{Discrete transformations in Fock space}
\label{sec:discrete}

We now turn our attention in this section to the definition of the discrete symmetry transformations of these non-Hermitian quantum field theories in Fock space. In particular, we define the $\hat{\mathcal{C}}'$ operator, and show that the parity and time-reversal operators are uniquely defined, irrespective of the choice of inner product.

\subsection{Parity}
\label{sec:parity}

We begin with the parity transformation, under which
the spatial coordinates $\mathbf{x}$ change sign, i.e., $\mathbf{x}\to \mathbf{x}'=-\mathbf{x}$, but not the time coordinate $t$, so that
\begin{equation}
x^{\mu} \equiv (t,\mathbf{x})\to \mathcal{P}x^{\mu}=x^{\prime\mu}= (t^{\prime},\mathbf{x}^{\prime})=(t,-\mathbf{x})~.
\end{equation}
A $c$-number complex scalar field transforms under parity as
\begin{equation}
\label{eq:cKGparity}
\mathcal{P}:\ \phi(x)\to \phi'(x')=\phi'(t,-\mathbf{x})=\eta_{\mathcal{P}}\phi(t,\mathbf{x})~,
\end{equation}
where $\eta_{\mathcal{P}}$ satisfies $|\eta_{\mathcal{P}}|^2=1$. If $\phi=\phi^*$ is real then $\eta_{\mathcal{P}}$ is equal to $+1$ if $\phi$ transforms as a scalar and equal to $-1$ if $\phi$ transforms as a pseudoscalar.~\footnote{It is always possible to rephase the parity operator such that spin-0 fields transform up to a real-valued phase of $\pm 1$, as we assume here.}

Requiring that the matrix elements of the quantum field operator $\hat{\phi}_i$ transform as in Eq.~\eqref{eq:revisedP} [see also Eq.~\eqref{eq:cKGparity}], we obtain
{\begin{subequations}
\label{eq:HermPphi}
\begin{align}
\hat{\mathcal{P}}\hat{\phi}_i(x)\hat{\mathcal{P}}^{-1}&=P_{ij}\check{\phi}_j(\mathcal{P}x)~,\\
\hat{\mathcal{P}}\hat{\phi}^{\dag}_i(x)\hat{\mathcal{P}}^{-1}&=P_{ij}\check{\phi}^{\dag}_j(\mathcal{P}x)~,
\end{align}
\end{subequations}
which are consistent with Eq.~\eqref{eq:phivarrels}.} As we show below, the definition of $\hat{\mathcal{P}}$ and its action on the field operators do not depend on the choice of inner product that defines the matrix elements. In terms of these creation and annihilation operators, the parity operator has the following explicit form~\cite{GreinerReinhardt}:
\begin{equation}
    \label{eq:scalParop}
    \hat{\mathcal{P}}=\exp\left\{\frac{i\pi}{2}\int_{\mathbf{p}}\left[\hat{a}^{\dag}_{i,\mathbf{p}}(0)\hat{a}_{i,-\mathbf{p}}(0)+\hat{c}^{\dag}_{i,\mathbf{p}}(0)\hat{c}_{i,-\mathbf{p}}(0)-\hat{a}^{\dag}_{i,\mathbf{p}}(0)P_{ij}\hat{a}_{j,\mathbf{p}}(0)-\hat{c}^{\dag}_{i,\mathbf{p}}(0)P_{ij}\hat{c}_{i,\mathbf{p}}(0)\right]\right\}~.
\end{equation}
We note that this operator is time independent, and can therefore be written in terms of Hermitian-conjugate creation and annihilation operators at the time $t=0$.

\subsection{$\mathcal{C}'$ Transformation}
\label{sec:Cprime}

Using the $Q$ matrix of the simplified model in Sec.~\ref{sec:bosonicmodel}, it is straightforward to construct the $\hat{\mathcal{C}}'$ operator for the model, which is given by
\begin{equation}
    \label{eq:Cprimenew}
    \hat{\mathcal{C}}'=\exp\left[{\rm arctanh}\,\eta \int_{\mathbf{p}}\left(\hat{a}_{i,\mathbf{p}}^{\dag}(0)\bar{Q}_{ij}\hat{a}_{j,\mathbf{p}}(0)-\hat{c}_{i,\mathbf{p}}^{\dag}(0)\bar{Q}_{ij}\hat{c}_{j,\mathbf{p}}(0)\right)\right]\hat{\mathcal{P}}_+\hat{\mathcal{P}}~,
\end{equation}
where the matrix $\bar{Q}$ is given in the flavour basis in Eq.~\eqref{eq:barQdef}. The relative sign between the bracketed particle and antiparticle operator terms in the exponent of Eq.~\eqref{eq:Cprimenew} ensures that the field operators transform appropriately, and reflects the fact that particle and antiparticle states must transform in the opposite sense (see below). We point out that the $\hat{\mathcal{C}}'$ operator is ill-defined at the exceptional point $\eta=1$, as is expected for this operator. Comparing with Eq.~\eqref{eq:matrixCprimeidentity}, we note the necessity of including an additional operator
\begin{equation}
    \label{eq:Pplus}
    \hat{\mathcal{P}}_+=\exp\left\{\frac{i\pi}{2}\int_{\mathbf{p}}\left[\hat{a}^{\dag}_{i,\mathbf{p}}(0)\hat{a}_{i,-\mathbf{p}}(0)+\hat{c}^{\dag}_{i,\mathbf{p}}(0)\hat{c}_{i,-\mathbf{p}}(0)-\hat{a}^{\dag}_{i,\mathbf{p}}(0)\hat{a}_{i,\mathbf{p}}(0)-\hat{c}^{\dag}_{i,\mathbf{p}}(0)\hat{c}_{i,\mathbf{p}}(0)\right]\right\}~,
\end{equation}
which implements the correct change of sign of the momentum in the $\mathcal{C}'\mathcal{PT}$ inner product. For transformations in Fock space, the $\hat{\mathcal{C}}'$ operator can be written in the form
\begin{equation}
    \label{eq:operatorCprimedef}
    \hat{\mathcal{C}}'=e^{-\hat{\mathcal{Q}}}\hat{\mathcal{P}}_+\hat{\mathcal{P}}~,
\end{equation}
where the operator $\hat{\mathcal{Q}}$ is discussed below.

In terms of the canonically conjugate field variables, the $\hat{\mathcal{C}}'$ operator can be written in the form
\begin{equation}
    \label{eq:Cprimefields}
    \hat{\mathcal{C}}'=\exp\left[-i\,{\rm arctanh}\,\eta \int_{\mathbf{x}}\left(\hat{\pi}_{i}(t,\mathbf{x})\bar{Q}_{ij}\hat{\phi}_{j}(t,\mathbf{x})-\check{\pi}^{\dag}_{i}(t,\mathbf{x})\bar{Q}_{ij}\check{\phi}^{\dag}_{j}(t,\mathbf{x})\right)\right]\hat{\mathcal{P}}_+\hat{\mathcal{P}}~.
\end{equation}
We draw attention to the appearance of both hatted and checked operators, cf.~Subsec.~\ref{flavour} and the canonical algebra in Eqs.~\eqref{eq:hatCCR} and~\eqref{eq:checkCCR}.

We emphasize that the $\hat{\mathcal{C}}'$ operator does not coincide with the usual charge-conjugation operator, which is~\cite{GreinerReinhardt}
\begin{equation}
     \hat{\mathcal{C}}=\exp\left\{\frac{i\pi}{2}\int_{\mathbf{p}}\left[\hat{c}^{\dag}_{i,\mathbf{p}}(0)\hat{a}_{i,-\mathbf{p}}(0)+\hat{a}^{\dag}_{i,\mathbf{p}}(0)\hat{c}_{i,-\mathbf{p}}(0)-\left(\hat{a}^{\dag}_{i,\mathbf{p}}(0){C_{ij}}\hat{a}_{j,\mathbf{p}}(0)+\hat{c}^{\dag}_{i,\mathbf{p}}(0){C_{ij}}\hat{c}_{j,\mathbf{p}}(0)\right)\right]\right\}\,.
\end{equation}
The charge matrix $C_{ij}$ must be chosen such that $C_{ij}=P_{ij}$ in order for the Lagrangian to be $\mathcal{C}$ symmetric, as a result of which $\hat{\mathcal{C}}$ and $\hat{\mathcal{C}}'$ do not commute. We note that the $\hat{\mathcal{C}}'$ operator depends on the non-Hermitian parameter $\eta$, whereas the usual charge-conjugation operator $\hat{\mathcal{C}}$ does not.

The action of $\hat{\mathcal{C}}'$ is as follows:
\begin{subequations}
\label{eq:scalCprimeaadag}
 \begin{align}
     \hat{\mathcal{C}}'\hat{a}^{\dag}_{i,\mathbf{q}}(0)\hat{\mathcal{C}}^{\prime,-1}=C^{\prime}_{ij}\hat{a}^{\dag}_{j,\mathbf{q}}(0)~,\\
     \hat{\mathcal{C}}'\hat{a}_{i,\mathbf{q}}(0)\hat{\mathcal{C}}^{\prime,-1}=C^{\prime\mathsf{T}}_{ij}\hat{a}_{j,\mathbf{q}}(0)~,\\
     \hat{\mathcal{C}}'\hat{c}^{\dag}_{i,\mathbf{q}}(0)\hat{\mathcal{C}}^{\prime,-1}=C^{\prime\mathsf{T}}_{ij}\hat{c}^{\dag}_{j,\mathbf{q}}(0)~,\\
     \hat{\mathcal{C}}'\hat{c}_{i,\mathbf{q}}(0)\hat{\mathcal{C}}^{\prime,-1}=C^{\prime}_{ij}\hat{c}_{j,\mathbf{q}}(0)~,
 \end{align}
\end{subequations}
with the fields transforming as
\begin{subequations}
\begin{gather}
    \hat{\mathcal{C}}'\hat{\phi}_{i}(x)\hat{\mathcal{C}}^{\prime,-1}=C^{\prime\mathsf{T}}_{ij}\hat{\phi}_{j}(x)~,\\
     \hat{\mathcal{C}}'\check{\phi}^{\dag}_{i}(x)\hat{\mathcal{C}}^{\prime,-1}=C^{\prime}_{ij}\check{\phi}^{\dag}_{j}(x) ~,
\end{gather}
\end{subequations}
such that
\begin{subequations}
 \begin{align}
     \hat{\phi}_1(x)\to \frac{1}{\sqrt{1-\eta^2}}\left(\hat{\phi}_1(x)+\eta \hat{\phi}_2(x)\right)~,
\quad \hat{\phi}_2(x)\to -\frac{1}{\sqrt{1-\eta^2}}\left(\hat{\phi}_2(x)+\eta \hat{\phi}_1(x)\right)~,\\
     \check{\phi}_1^{\dag}(x)\to \frac{1}{\sqrt{1-\eta^2}}\left(\check{\phi}_1^{\dag}(x)-\eta \check{\phi}_2^{\dag}(x)\right)~,
\quad \check{\phi}_2^{\dag}(x)\to -\frac{1}{\sqrt{1-\eta^2}}\left(\check{\phi}_2^{\dag}(x)-\eta \check{\phi}_1^{\dag}(x)\right)~.
 \end{align}
\end{subequations}
That $\hat{a}$ and $\hat{c}$ transform differently follows directly from the fact that $\hat{\mathcal{C}}'$ and the usual charge conjugation operator $\hat{\mathcal{C}}$ do not commute. It is easy to confirm that $\hat{\mathcal{C}}^{\prime 2}=\mathbb{I}$, and that $\mathcal{C}'$- and $\mathcal{PT}$-conjugation commute.~\footnote{We reiterate that $\mathcal{PT}$ conjugation, denoted here by $\ddagger=\mathcal{PT}\circ\mathsf{T}$, includes operator/matrix transposition.} To see this, consider the superposition of single-particle momentum states
\begin{equation} \ket{\Psi}=\int_{\mathbf{p}}A_{\mathbf{p},i}\ket{\mathbf{p}_i}~,
\end{equation}
where the $A_{\mathbf{p},i}$ are complex $c$-number coefficients. Acting first with $\ddagger$ and then with $\hat{\mathcal{C}}'$, and making use of Eq.~\eqref{eq:scalCprimeaadag}, we have
\begin{equation}
    \ket{\Psi}^{\ddagger}\hat{\mathcal{C}}'=\int_{\mathbf{p}}\bra{\mathbf{p}_j}P_{ji}A^*_{\mathbf{p},i}\hat{\mathcal{C}}'=\int_{\mathbf{p}}\bra{\mathbf{p}_k}C^{\prime}_{kj}P_{ji}A^*_{\mathbf{p},i}~.
\end{equation}
Conversely, we have
\begin{equation}
    \left(\hat{\mathcal{C}}'\ket{\Psi}\right)^{\ddagger}=\int_{\mathbf{p}}\left(A_{\mathbf{p},i}C'_{ij}\ket{\mathbf{p}_j}\right)^{\ddagger}=\int_{\mathbf{p}}\bra{\mathbf{p}_k}P_{kj}C^{\prime\mathsf{T}}_{ji}A^*_{\mathbf{p},i}~.
\end{equation}
Using the fact that
\begin{equation}
    P.C'.P=C^{\prime\mathsf{T}}~,
\end{equation}
we see that $\mathcal{C}'$- and $\mathcal{PT}$-conjugation commute, as required. Moreover, the Hamiltonian given by Eq.~\eqref{eq:hatHamiltonian} (and the corresponding Lagrangian) is $\mathcal{C}'$ symmetric, such that $[\hat{\mathcal{C}}',\hat{H}]=0$. Since the $\mathcal{C}'$ transformation mixes the scalar and pseudoscalar operators, we find that $\hat{\mathcal{C}}'$ does not commute with $\hat{\mathcal{P}}$. This is, in fact, a necessary consequence of the relation $\hat{\mathcal{P}}\hat{\mathcal{Q}}\hat{\mathcal{P}}=-\hat{\mathcal{Q}}$, as we discuss in the next Section.

\subsection{The Similarity Transformation}
\label{sec:similarity}

The $\hat{\mathcal{Q}}$ operator in Eq.~(\ref{eq:operatorCprimedef}) is given by
\begin{equation}
    \label{eq:Qdef}
    \hat{\mathcal{Q}}=-{\rm arctanh}\,\eta \int_{\mathbf{p}}\left(\hat{a}_{1,\mathbf{p}}^{\dag}(0)\hat{a}_{2,\mathbf{p}}(0)+\hat{a}_{2,\mathbf{p}}^{\dag}(0)\hat{a}_{1,\mathbf{p}}(0)-\hat{c}_{1,\mathbf{p}}^{\dag}(0)\hat{c}_{2,\mathbf{p}}(0)-\hat{c}_{2,\mathbf{p}}^{\dag}(0)\hat{c}_{1,\mathbf{p}}(0)\right)~.
\end{equation}
The similarity transformation $\hat{\mathcal{O}}\to e^{-\hat{\mathcal{Q}}/2}\hat{\mathcal{O}}e^{\hat{\mathcal{Q}}/2}$ has the following
effects on the particle and antiparticle annihilation and creation operators:
\begin{subequations}
 \begin{align}
     \hat{a}_{i,\mathbf{q}}(0)\to \hat{a}_{i,\mathbf{q}}(0)\cosh\frac{{\rm arctanh}\,\eta}{2}-\hat{a}_{\slashed{i},\mathbf{q}}(0)\sinh\frac{{\rm arctanh}\,\eta}{2}~,\\
     \hat{a}^{\dag}_{i,\mathbf{q}}(0)\to \hat{a}^{\dag}_{i,\mathbf{q}}(0)\cosh\frac{{\rm arctanh}\,\eta}{2}+\hat{a}^{\dag}_{\slashed{i},\mathbf{q}}(0)\sinh\frac{{\rm arctanh}\,\eta}{2}~,\\
     \hat{c}_{i,\mathbf{q}}(0)\to \hat{c}_{i,\mathbf{q}}(0)\cosh\frac{{\rm arctanh}\,\eta}{2}+\hat{c}_{\slashed{i},\mathbf{q}}(0)\sinh\frac{{\rm arctanh}\,\eta}{2}~,\\
     \hat{c}^{\dag}_{i,\mathbf{q}}(0)\to \hat{c}^{\dag}_{i,\mathbf{q}}(0)\cosh\frac{{\rm arctanh}\,\eta}{2}-\hat{c}^{\dag}_{\slashed{i},\mathbf{q}}(0)\sinh\frac{{\rm arctanh}\,\eta}{2}~,
 \end{align}
\end{subequations}
so that the fields transform as
\begin{subequations}
\begin{align}
    \hat{\phi}_i(x)\to \hat{\xi}_i(x)\cosh\frac{{\rm arctanh}\,\eta}{2}-\hat{\xi}_{\slashed{i}}(x)\sinh\frac{{\rm arctanh}\,\eta}{2}~,\\
    \check{\phi}_i^{\dag}(x)\to \hat{\xi}_i^{\dag}(x)\cosh\frac{{\rm arctanh}\,\eta}{2}+\hat{\xi}_{\slashed{i}}^{\dag}(x)\sinh\frac{{\rm arctanh}\,\eta}{2}~,
\end{align}
\end{subequations}
where $\hat{\xi}_i$ are the field operators in the mass eigenbasis. Herein, $\slashed{i}=2$ for $i=1$, and $\slashed{i}=1$ for $i=2$. Using
\begin{subequations}
 \begin{align}
 \hspace{-2mm}
     \cosh\frac{{\rm arctanh}\,\eta}{2}&=\frac{1}{\sqrt{2}}\sqrt{1+\frac{1}{\sqrt{1-\eta^2}}}~,\\
     \sinh\frac{{\rm arctanh}\,\eta}{2}&=\frac{1}{\sqrt{2}}\frac{\eta}{\sqrt{1-\eta^2}}\frac{1}{\sqrt{1+\frac{1}{\sqrt{1-\eta^2}}}}~,
 \end{align}
\end{subequations}
one can show with some algebra that this indeed gives the correct transformation to the Hermitian theory whose Lagrangian is~\footnote{Note that both the kinetic terms have positive signs, unlike in Ref.~\cite{Mannheim} (see also the Appendix).}
\begin{equation}
\label{eq:similarityLag}
    \hat{\mathcal{L}}'=\partial_{\nu}\hat{\xi}_1^{\dag}(x)\partial^{\nu}\hat{\xi}_1(x)+\partial_{\nu}\hat{\xi}_2^{\dag}(x)\partial^{\nu}\hat{\xi}_2(x)-m_+^2\hat{\xi}_1^{\dag}(x)\hat{\xi}_1(x)-m_-^2\hat{\xi}_2^{\dag}(x)\hat{\xi}_2(x)~.
\end{equation}
Note that the similarity-transformed Lagrangian is isospectral to the original Lagrangian. Hence, the \emph{non-interacting} non-Hermitian bosonic model is equivalent to a Hermitian theory.

We draw attention to the fact that we have used the form of the $\hat{\mathcal{Q}}$ operator extracted from Eq.~\eqref{eq:Cprimenew} in terms of the creation and annihilation operators evaluated at the initial time surface, and not from Eq.~\eqref{eq:Cprimefields} in terms of the field operators at the finite time $t$.  While both forms of the $\hat{\mathcal{Q}}$ operator give valid $\hat{\mathcal{C}}'$ transformations, only the former choice, in terms of the creation and annihilation operators, gives a similarity transformation that both maps the Lagrangian to the Hermitian one \emph{and} transforms the field operators to those of the mass eigenbasis.  Were we to take a $\hat{\mathcal{Q}}$ operator based on Eq.~\eqref{eq:Cprimefields}, it would map the Lagrangian to the Hermitian one, but leave the field operators themselves unchanged.  This would therefore not represent a consistent similarity transformation to the Hermitian ``frame''.  The reason for this discrepancy is the fact that, by virtue of the non-Hermitian nature of the evolution, the $\hat{\mathcal{C}}'$ operators in Eqs.~\eqref{eq:Cprimenew} and~\eqref{eq:Cprimefields} are actually distinct.
 
We have constructed a similarity transformation that maps the non-Hermitian free theory to a Hermitian one. If we include interactions, however, it is not in general the case that this similarity transformation will map the full interacting Hamiltonian to a Hermitian one, even if those interactions respect the $\mathcal{PT}$ symmetry. For example, if one adds a {Hermitian} quartic interaction term $\lambda \left(\phi_1^{\dag}\phi_1\right)^2$ to the non-Hermitian bosonic model, as
 discussed in the context of spontaneous symmetry breaking in Refs.~\cite{AEMS1, AEMS2, AEMS3}, the
 similarity transformation converts it into a
 non-Hermitian combination of $\xi_1$, $\xi_2$, $\xi_1^{\dag}$ and $\xi_2^{\dag}$:
\begin{align}
\label{eq:quartic}
    \frac{\lambda}{4}\left(\check{\phi}_1^{\dag}\hat{\phi}_1\right)^2&\to\frac{\lambda}{16}\left[\left(\frac{m_1^2-m_2^2}{2}\right)\left(m_+^2-m_2^2\right)-\mu^4\right]^{-2} \nonumber\\&\times\left[\left(m_+^2-m_2^2\right)\hat{\xi}_1^{\dag}+\mu^2\hat{\xi}_2^{\dag}\right]^2\left[\left(m_+^2-m_2^2\right)\hat{\xi}_1-\mu^2\hat{\xi}_2\right]^2~.
\end{align}
Hence, this \emph{interacting} non-Hermitian bosonic model is \emph{not} equivalent to a Hermitian theory according to the above similarity transformation. Instead,
it exhibits soft breaking of Hermiticity. On the other hand, were we to build interaction terms out of the quadratic field invariants discussed in Subsec.~\ref{sec:analogy}, or powers of the mass term, the above similarity transformation would map these to Hermitian interactions.

This observation does not necessarily indicate that the eigenvalues of the interacting Hamiltonian are complex or preclude the possibility that there exists a different similarity transformation that maps the full interacting Hamiltonian to a Hermitian one. It does, however, present a challenge for the perturbative treatment of non-Hermitian theories, since the interaction pictures of the non-Hermitian and corresponding Hermitian theories would necessarily have to be related by a similarity transformation that would involve a resummation of a series in the coupling constant that may be non-trivial. We leave further study of this interesting point to future work.

Finally, we comment on the connection of this similarity transformation to the $\mathcal{V}$ norm considered in Ref.~\cite{Mannheim:2017apd} at the level of the free theory. The $\mathcal{V}$ norm is constructed from the operator $\hat{\mathcal{V}}=e^{-\hat{\mathcal{Q}}}$, which maps the Hamiltonian to its Hermitian conjugate, i.e., $\hat{H}^{\dag}=\hat{\mathcal{V}}\hat{H}\hat{\mathcal{V}}^{-1}$. Since we also have that $\hat{H}^{\dag}=\hat{\mathcal{P}}\hat{H}\hat{\mathcal{P}}^{-1}$, and $\hat{\mathcal{P}}_+\hat{H}\hat{\mathcal{P}}_+^{-1}=\hat{H}$ it follows that $[\hat{\mathcal{V}}\hat{\mathcal{P}}_+\hat{\mathcal{P}},\hat{H}]=0$, and we can identify $\hat{\mathcal{C}}'=\hat{\mathcal{V}}\hat{\mathcal{P}}_+\hat{\mathcal{P}}$, so long as $(\hat{\mathcal{V}}\hat{\mathcal{P}}_+\hat{\mathcal{P}})^2=\mathbb{I}$. This is indeed the case, since $\hat{\mathcal{P}}_+\hat{\mathcal{Q}}\hat{\mathcal{P}}_+=\hat{\mathcal{Q}}$ and $\hat{\mathcal{P}}\hat{\mathcal{Q}}\hat{\mathcal{P}}=-\hat{\mathcal{Q}}$. The latter identity follows immediately from Eq.~\eqref{eq:Qdef}, upon realising that $\hat{\mathcal{Q}}$ is bilinear in the scalar and pseudoscalar operators. (We recall that $\hat{\mathcal{P}}$ and $\hat{\mathcal{P}}_+$ are involutary). Since the $\hat{\mathcal{V}}$ norm of Ref.~\cite{Mannheim:2017apd} is positive definite by construction, the same follows for the $\hat{\mathcal{C}'}\hat{\mathcal{P}}\hat{\mathcal{T}}$ norm constructed in this work, since they coincide, as we will show below. An explicit comparison of the $\mathcal{C}'\mathcal{P}\mathcal{T}$ and $\mathcal{V}$ norms in the case of interacting theories warrants further investigation beyond the scope of this article.

\subsection{Inner products}
\label{sec:products}

Before we can consider the definition of the time-reversal operator in Fock space, we must first describe the various inner products with respect to which it can be defined.  For this purpose, it is convenient to define a 
variation of Dirac's bra-ket notation in which the bra and ket states are related by transposition rather than Hermitian conjugation. Specifically, we define
\begin{equation}
\label{eq:branotation}
\bra{\alpha}\equiv (\ket{\alpha})^{\mathsf{T}}~,
\end{equation}
where $\mathsf{T}$ denotes transposition. Hermitian conjugation is indicated in the usual way by a superscript $\dag$ denoting the
combination $\dag\equiv \ast\circ\mathsf{T}$, where $\ast$ indicates complex conjugation.

We can now distinguish the following inner products in Fock space: 

\begin{listcases}

\item {\bf Dirac inner product:} In this notation, the usual Dirac inner product, which is defined via Hermitian conjugation, is written as
\begin{equation}
(\ket{\alpha})^{\dag}\ket{\beta}=\braket{\alpha^*|\beta}=\braket{\hat{\mathcal{K}}^{\mathsf{T}}\alpha|\beta}=\braket{\alpha|\hat{\mathcal{K}}|\beta}=\braket{\alpha|\hat{\mathcal{K}}\beta}~,
\end{equation}
where the antilinear operator $\hat{\mathcal{K}}$ is $\propto \hat{\mathcal{T}}$ and effects complex conjugation. For a spin-zero field, single-particle states of momentum $\mathbf{q}$ and $\mathbf{q}'$ have the usual Dirac normalization
\begin{equation}
    (\ket{\mathbf{q}})^{\dag}\ket{\mathbf{q}'}=\braket{\mathbf{q}|\mathbf{q}'}=(2\pi)^3\delta^{3}(\mathbf{q}-\mathbf{q}')~.
\end{equation}

\item {\bf $\mathcal{PT}$ inner product:} This indefinite
inner product is defined via $\mathcal{PT}$ conjugation, 
which we denote by $\ddag\equiv \mathcal{PT}\circ \mathsf{T}$, and is written as
\begin{equation}
(\ket{\alpha})^{\ddag}\ket{\beta}=\braket{\alpha^{\mathcal{PT}}|\beta}=\braket{\hat{\mathcal{T}}^{\mathsf{T}}\hat{\mathcal{P}}^{\mathsf{T}}\alpha|\beta}=\braket{\alpha|\hat{\mathcal{P}}\hat{\mathcal{T}}|\beta}=\braket{\alpha|\hat{\mathcal{P}}\hat{\mathcal{T}}\beta}~.
\end{equation}
For a scalar field, the $\mathcal{PT}$ inner product of single-particle momentum eigenstates is
\begin{equation}
    (\ket{\mathbf{q}})^{\ddag}\ket{\mathbf{q}'}=\braket{\hat{\mathcal{T}}^{\mathsf{T}}\hat{\mathcal{P}}^{\mathsf{T}}\mathbf{q}|\mathbf{q}'}=\eta_{\mathcal{P}}\braket{\mathbf{q}|\mathbf{q}}=\eta_{\mathcal{P}}(2\pi)^3\delta^3(\mathbf{q}-\mathbf{q}')~,
\end{equation}
which is negative definite in the case of a pseudoscalar ($\eta_{\mathcal{P}}=-1$), cf.~the approach of Ref.~\cite{Mannheim:2017apd}.

\item {\bf $\mathcal{C'PT}$ inner product:} This positive-definite {inner product} is defined via $\mathcal{C'PT}$ conjugation, which we denote by $\S\equiv \mathcal{C'PT}\circ \mathsf{T}$, and is written as
\begin{equation}
(\ket{\alpha})^{\S}\ket{\beta}=\braket{\alpha^{\mathcal{C'PT}}|\beta}=\braket{\hat{\mathcal{T}}^{\mathsf{T}}\hat{\mathcal{P}}^{\mathsf{T}}\hat{\mathcal{C}}^{\prime\mathsf{T}}\alpha|\beta}=\braket{\alpha|\hat{\mathcal{C}}'\hat{\mathcal{P}}\hat{\mathcal{T}}|\beta}=\braket{\alpha|\hat{\mathcal{C}}'\hat{\mathcal{P}}\hat{\mathcal{T}}\beta}~.
\end{equation}
With respect to this inner product, the norm of the single-particle momentum state is positive definite for both the scalar and pseudoscalar:
\begin{equation}
        (\ket{\mathbf{q}})^{\S}\ket{\mathbf{q}}=\braket{\hat{\mathcal{T}}^{\mathsf{T}}\hat{\mathcal{P}}^{\mathsf{T}}\hat{\mathcal{C}}^{\prime\mathsf{T}}\mathbf{q}|\mathbf{q}}=\eta_{\mathcal{P}}\braket{\hat{\mathcal{T}}^{\mathsf{T}}\hat{\mathcal{P}}^{\mathsf{T}}\mathbf{q}|\mathbf{q}}=\eta_{\mathcal{P}}^2\braket{\mathbf{q}|\mathbf{q}}=1~.
\end{equation}
Here, we have simply taken $\eta\to 0$ in Eqs.~\eqref{eq:Cprimenew} and \eqref{eq:scalCprimeaadag} in order to decouple the flavours. In this case, $\hat{\phi}^{\S}(x)=\hat{\phi}^{\dag}(x)$, trivially, i.e., in the Hermitian limit $\eta\to 0$, $C'\mathcal{PT}$ conjugation of the field operator coincides with Hermitian conjugation.

Returning to the two-flavour case, we can take the single-particle mass eigenstate
\begin{equation}
    \ket{\mathbf{p}_+}=N\left\{\eta\ket{\mathbf{p}_1}+\left[1-\sqrt{1-\eta^2}\right]\ket{\mathbf{p}_2}\right\}
\end{equation}
as an example. By our notation in Eq.~\eqref{eq:branotation}, we have
\begin{equation}
    \bra{\mathbf{p}_+}=N\left\{\eta\bra{\mathbf{p}_1}+\left[1-\sqrt{1-\eta^2}\right]\bra{\mathbf{p}_2}\right\}~.
\end{equation}
In addition, it follows from the action of the $\hat{\mathcal{C}}'$ and $\hat{\mathcal{P}}\hat{\mathcal{T}}$ operators that
\begin{equation}
    \hat{\mathcal{C}}'\hat{\mathcal{P}}\hat{\mathcal{T}}\ket{\mathbf{p}_+}=N\left\{\eta\ket{\mathbf{p}_1}-\left[1-\sqrt{1-\eta^2}\right]\ket{\mathbf{p}_2}\right\}~.
\end{equation}
It is then easily verified that
\begin{equation}
    \braket{\mathbf{p}_+|\hat{\mathcal{C}}'\hat{\mathcal{P}}\hat{\mathcal{T}}|\mathbf{p}_+}=1>0~,
\end{equation}
as required. Moreover, we have that
\begin{equation}
    \hat{\mathcal{P}}_+\hat{\mathcal{T}}\ket{\mathbf{p}_+}=N\left\{\eta\ket{\mathbf{p}_1}+\left[1-\sqrt{1-\eta^2}\right]\ket{\mathbf{p}_2}\right\}=\ket{\mathbf{p}_+}~.
\end{equation}
We can then confirm, as highlighted earlier, that the $\mathcal{C}'\mathcal{P}\mathcal{T}$ and $\mathcal{V}$ norms coincide for the free theory:
\begin{equation}
    \braket{\mathbf{p}_+|\hat{\mathcal{C}}'\hat{\mathcal{P}}\hat{\mathcal{T}}|\mathbf{p}_+}=\braket{\mathbf{p}_+|\hat{\mathcal{V}}\hat{\mathcal{P}}_+\hat{\mathcal{P}}^2\hat{\mathcal{T}}|\mathbf{p}_+}=\braket{\mathbf{p}_+|\hat{\mathcal{V}}\hat{\mathcal{P}}_+\hat{\mathcal{T}}|\mathbf{p}_+}=\braket{\mathbf{p}_+|\hat{\mathcal{V}}|\mathbf{p}_+}~.
\end{equation}

\end{listcases}

\subsection{Parity Revisited}
\label{sec:parity2}

Having defined the various inner products, we can now return to the parity operator, and show explicitly that its definition does not depend on which inner product we use to construct the matrix elements of the theory.

\begin{listcases}

\item {\bf Dirac inner product:} In this case, the transformation rules for the ket and bra states are
\begin{equation}
\ket{\hat{\mathcal{P}}\alpha}=\hat{\mathcal{P}}\ket{\alpha}\qquad \Leftrightarrow \qquad (\ket{\hat{\mathcal{P}}\alpha})^{\dag}=(\hat{\mathcal{P}}\ket{\alpha})^{\dag}=\bra{\alpha^*}\hat{\mathcal{P}}^{\dag}=\bra{\alpha^*}\hat{\mathcal{P}}^{-1}~.
\end{equation}
We note that parity and Hermitian conjugation commute, so that
\begin{equation}
\braket{(\hat{\mathcal{P}}^{\mathsf{T}}\alpha)^*|\check{\phi}_i(\mathcal{P}x)|\hat{\mathcal{P}}\beta}=\braket{\alpha^*|\hat{\mathcal{P}}^{-1}\check{\phi}_i(\mathcal{P}x)\hat{\mathcal{P}}|\beta}\overset{!}{=}P_{ij}\braket{\alpha^*|\hat{\phi}_j(x)|\beta}~,
\end{equation}
and we recover the results in Eq.~\eqref{eq:HermPphi}.

\item {\bf $\mathcal{PT}$ inner product:} The situation is similar in this case, because $\hat{\mathcal{P}}$ and $\hat{\mathcal{T}}$ commute (so long as $\eta_{\mathcal{P}}\in\mathbb{R}$). Specifically, the transformation rules for the ket and bra states are
\begin{equation}
\ket{\hat{\mathcal{P}}\alpha}=\hat{\mathcal{P}}\ket{\alpha}\qquad \Leftrightarrow \qquad (\ket{\hat{\mathcal{P}}\alpha})^{\ddag}=(\hat{\mathcal{P}}\ket{\alpha})^{\ddag}=\bra{\alpha^{\mathcal{PT}}}\hat{\mathcal{P}}^{\ddag}=\bra{\alpha^{\mathcal{PT}}}\hat{\mathcal{P}}^{-1}~,
\end{equation}
where $\hat{\mathcal{P}}^{\ddag}=(\hat{\mathcal{P}}\hat{\mathcal{T}})\hat{\mathcal{P}}^{\mathsf{T}}(\hat{\mathcal{T}}^{-1}\hat{\mathcal{P}}^{-1})$. We therefore recover the same transformation rules~\eqref{eq:HermPphi} for the field operators as in the Hermitian case.
This is perhaps not surprising, since Hermitian conjugation is substituted by $\mathcal{PT}$ conjugation in non-Hermitian theories.

\item {\bf $\mathcal{C'PT}$ inner product:}  This case is rather different, since the $\mathcal{C}'$ and $\mathcal{P}$ transformations do not commute. The transformation rules for the ket and bra states are therefore
\begin{subequations}
\begin{align}
\ket{\hat{\mathcal{P}}\alpha}=\hat{\mathcal{P}}\ket{\alpha}\qquad &\Leftrightarrow \qquad (\ket{\hat{\mathcal{P}}\alpha})^{\S}=(\hat{\mathcal{P}}\ket{\alpha})^{\S}=\bra{\alpha^{\mathcal{C'PT}}}\hat{\mathcal{P}}^{\S}=\bra{\alpha^{\mathcal{C'PT}}}\hat{\mathcal{C}}'\hat{\mathcal{P}}\hat{\mathcal{C}}'\\
&\Leftrightarrow \qquad (\ket{\alpha})^{\S}\hat{\mathcal{P}}^{\mathsf{T}}=(\hat{\mathcal{P}}\ket{\alpha^{\mathcal{C}'\mathcal{PT}}})^{\mathsf{T}}=\bra{\alpha^{\mathcal{C}'\mathcal{PT}}}\hat{\mathcal{P}}^{-1}~.
\end{align}
\end{subequations}
It is the matrix element involving the latter that leads to a definition of the parity operator consistent with Eq.~\eqref{eq:scalParop}, and we then have
\begin{equation}
\braket{\hat{\mathcal{P}}^{\mathsf{T}}(\alpha^{\mathcal{C'PT}})|\check{\phi}_i(\mathcal{P}x)|\hat{\mathcal{P}}\beta}=\braket{\alpha^{\mathcal{C'PT}}|\hat{\mathcal{P}}^{-1}\check{\phi}_i(\mathcal{P}x)\hat{\mathcal{P}}|\beta}\overset{!}{=}P_{ij}\braket{\alpha^{\mathcal{C'PT}}|\hat{\phi}_j(x)|\beta}~,
\end{equation}
giving the same transformation rules~\eqref{eq:HermPphi}.

\end{listcases}
 
\subsection{Time Reversal}
\label{sec:time}

Under a time-reversal transformation, the time coordinate $t\to t'=-t$, and
\begin{equation}
x^{\mu}\equiv(t,\mathbf{x})\to \mathcal{T}x^{\mu}=x^{\prime\mu}=(t^{\prime},\mathbf{x}^{\prime})=(-t,\mathbf{x})~.
\end{equation}
In this case a $c$-number complex Klein-Gordon field transforms as
\begin{equation}
\mathcal{T}:\ \phi(x)\to \phi^{\prime}(x^{\prime})=\phi^{\prime}(-t,\mathbf{x})=\eta_{\mathcal{T}}\phi^*(t,\mathbf{x})~,
\end{equation}
where $|\eta_{\mathcal{T}}|^2=1$. When translating this transformation to the corresponding $q$-number field operator, we need to take into account
the fact that time reversal interchanges the initial and final states. It is for this reason that the action of the time-reversal operator on field operators depends on the inner product used to determine the matrix elements. {However, as we see below, the time-reversal operator remains uniquely defined.}

\begin{listcases}

\item {\bf Dirac inner product:} In the case of the Dirac inner product, the transformation rules for the ket and bra states are
\begin{equation}
\ket{\hat{\mathcal{T}}\alpha}=\hat{\mathcal{T}}\ket{\alpha}\qquad \Leftrightarrow \qquad (\ket{\hat{\mathcal{T}}\alpha})^{\dag}=(\hat{\mathcal{T}}\ket{\alpha})^{\dag}=\bra{\alpha^*}\hat{\mathcal{T}}^{\dag}=\bra{\alpha^*}\hat{\mathcal{T}}^{-1}~.
\end{equation}
We note that time reversal and Hermitian conjugation commute (for $T_{ij}\in \mathbb{R}$), so that
\begin{equation}
\braket{(\mathcal{T}^{\mathsf{T}}\alpha)^*|\hat{\phi}_i(\mathcal{T}x)|\hat{\mathcal{T}}\beta}=\braket{\alpha^*|\hat{\mathcal{T}}^{-1}\hat{\phi}_i(\mathcal{T}x)\hat{\mathcal{T}}|\beta}\overset{!}{=}T_{ij}\braket{\beta^*|\hat{\phi}_j^{\dag}(x)|\alpha}~.
\end{equation}
Making use of the following identity that holds for an antilinear operator:
\begin{equation}
\braket{\alpha^*|\hat{\mathcal{T}}^{-1}\hat{\phi}_i(\mathcal{T}x)\hat{\mathcal{T}}|\beta}=\braket{\beta^*|\big(\hat{\mathcal{T}}^{-1}\hat{\phi}_i(\mathcal{T}x)\hat{\mathcal{T}}\big)^{\dag}|\alpha}~,
\end{equation}
we arrive at the familiar transformations
\begin{subequations}
\label{eq:usualTphi}
\begin{align}
\hat{\mathcal{T}}\hat{\phi}_i(x)\hat{\mathcal{T}}^{-1}&=T_{ij}\hat{\phi}_j(\mathcal{T}x)~,\\
\hat{\mathcal{T}}\hat{\phi}_i^{\dag}(x)\hat{\mathcal{T}}^{-1}&=T_{ij}^*\hat{\phi}^{\dag}_j(\mathcal{T}x)~.
\end{align}
\end{subequations}
Choosing both the scalar and pseudoscalar of our prototype model to transform with a phase of $+1$ under time reversal, the explicit form of the time-reversal operator is (see, e.g., Ref.~\cite{BjorkenDrell})
\begin{equation}
    \hat{\mathcal{T}}=\hat{\mathcal{K}}\hat{\mathcal{P}}_+~,
\end{equation}
where $\hat{\mathcal{K}}$ is the operator that effects complex conjugation on $c$-numbers and $\hat{\mathcal{P}}_+$ is the operator defined in Eq.~\eqref{eq:Pplus}.

\item {\bf $\mathcal{PT}$ inner product:} For the $\mathcal{PT}$-conjugate states, the transformation rules for the ket and bra states are
\begin{equation}
\ket{\hat{\mathcal{T}}\alpha}=\hat{\mathcal{T}}\ket{\alpha}\qquad \Leftrightarrow \qquad (\ket{\hat{\mathcal{T}}\alpha})^{\ddag}=(\hat{\mathcal{T}}\ket{\alpha})^{\ddag}=\bra{\alpha^{\mathcal{PT}}}\hat{\mathcal{T}}^{\ddag}=\bra{\alpha^{\mathcal{PT}}}\hat{\mathcal{T}}^{-1}~,
\end{equation}
where we have used $\hat{\mathcal{T}}\hat{\mathcal{T}}^{\mathsf{T}}\hat{\mathcal{T}}^{-1}=\hat{\mathcal{T}}^{\dag}$. In this case, we have~\footnote{Taking $T_{ij}=\delta_{ij}$ for simplicity, the action of an antilinear operator on the $\mathcal{PT}$ inner product is
\begin{align}
    \braket{\alpha^{\mathcal{PT}}|\hat{\mathcal{T}}^{-1}\hat{\phi}_i(\mathcal{T}x)\hat{\mathcal{T}}|\beta}=\braket{\beta^*|\hat{\phi}^{\dag}_i(x)|\alpha^{\mathcal{PT}*}}=\braket{\beta^{\mathcal{PT}}|\hat{\mathcal{K}}\hat{\mathcal{P}}\hat{\mathcal{T}}\hat{\phi}^{\dag}_i(x)\hat{\mathcal{K}}\hat{\mathcal{P}}\hat{\mathcal{T}}|\alpha}=\braket{\beta^{\mathcal{PT}}|\hat{\phi}^{\ddag}_i(x)|\alpha}~. \nonumber
\end{align}}
\begin{equation}
\braket{(\hat{\mathcal{T}}^{\mathsf{T}}\alpha)^{\mathcal{PT}}|\hat{\phi}_i(\mathcal{T}x)|\hat{\mathcal{T}}\beta}=\braket{\alpha^{\mathcal{PT}}|\hat{\mathcal{T}}^{-1}\hat{\phi}_i(\mathcal{T}x)\hat{\mathcal{T}}|\beta}\overset{!}{=}T_{ij}\braket{\beta^{\mathcal{PT}}|\hat{\phi}_j^{\ddag}(x)|\alpha}~.
\end{equation}
Making use of the identity
\begin{equation}
\braket{\alpha^{\mathcal{PT}}|\hat{\mathcal{T}}^{-1}\hat{\phi}_i(\mathcal{T}x)\hat{\mathcal{T}}|\beta}=\braket{\beta^{\mathcal{PT}}|\big(\hat{\mathcal{T}}^{-1}\hat{\phi}_i(\mathcal{T}x)\hat{\mathcal{T}}\big)^{\ddag}|\alpha}~,
\end{equation}
we quickly recover the transformations in Eq.~\eqref{eq:usualTphi}.

\item {\bf $\mathcal{C'PT}$ inner product:} Without making any assumption as to whether the $\mathcal{C}'$ and $\mathcal{T}$ transformations commute, the transformation rules for the ket and bra states for the $\mathcal{C}'\mathcal{PT}$ inner product are
\begin{subequations}
\begin{align}
\label{eq:T1}
\ket{\hat{\mathcal{T}}\alpha}=\hat{\mathcal{T}}\ket{\alpha}\qquad &\Leftrightarrow \qquad (\ket{\hat{\mathcal{T}}\alpha})^{\S}=(\hat{\mathcal{T}}\ket{\alpha})^{\S}=\bra{\alpha^{\mathcal{C'PT}}}\hat{\mathcal{T}}^{\S}=\bra{\alpha^{\mathcal{C'PT}}}\hat{\mathcal{C}}'\hat{\mathcal{T}}\hat{\mathcal{C}}'\\
\label{eq:T2}
&\Leftrightarrow \qquad (\ket{\alpha})^{\S}\hat{\mathcal{T}}^{\mathsf{T}}=(\hat{\mathcal{T}}\ket{\alpha^{\mathcal{C}'\mathcal{PT}}})^{\mathsf{T}}=\bra{\alpha^{\mathcal{C'PT}}}\hat{\mathcal{T}}^{-1}~.
\end{align}
\end{subequations}
Taking matrix elements involving the latter, we require~\footnote{Taking $T_{ij}=\delta_{ij}$ for simplicity, the action of an antilinear operator on the $\mathcal{C}'\mathcal{PT}$ inner product is:
\begin{align}
    \braket{\alpha^{\mathcal{C}'\mathcal{PT}}|\hat{\mathcal{T}}^{-1}\hat{\phi}_i(\mathcal{T}x)\hat{\mathcal{T}}|\beta}=\braket{\beta^*|\hat{\phi}^{\dag}_i(x)|\alpha^{\mathcal{C}'\mathcal{PT}*}}=\braket{\beta^{\mathcal{C}'\mathcal{PT}}|\hat{\mathcal{K}}\mathcal{C}'\hat{\mathcal{P}}\hat{\mathcal{T}}\hat{\phi}^{\dag}_i(x)\hat{\mathcal{K}}\mathcal{C}'\hat{\mathcal{P}}\hat{\mathcal{T}}|\alpha}=\braket{\beta^{\mathcal{C}'\mathcal{PT}}|\hat{\phi}^{\S}_i(x)|\alpha}~. \nonumber
\end{align}}
\begin{equation}
\braket{\hat{\mathcal{T}}^{\mathsf{T}}(\alpha^{\mathcal{C}'\mathcal{PT}})|\hat{\phi}_i(\mathcal{T}x)|\hat{\mathcal{T}}\beta}=\braket{\alpha^{\mathcal{C}'\mathcal{PT}}|\hat{\mathcal{T}}^{-1}\hat{\phi}_i(\mathcal{T}x)\hat{\mathcal{T}}|\beta}\overset{!}{=}T_{ij}\braket{\beta^{\mathcal{C}'\mathcal{PT}}|\hat{\phi}^{\S} _j(x)|\alpha}~.
\end{equation}
Making use of the identity
\begin{equation}
\braket{\alpha^{\mathcal{C}'\mathcal{PT}}|\hat{\mathcal{T}}^{-1}\hat{\phi}_i(\mathcal{T}x)\hat{\mathcal{T}}|\beta}=\braket{\beta^{\mathcal{C}'\mathcal{PT}}|\big(\hat{\mathcal{T}}^{-1}\hat{\phi}_i(\mathcal{T}x)\hat{\mathcal{T}}\big)^{\S}|\alpha}~,
\end{equation}
and we again recover the transformations in Eq.~\eqref{eq:usualTphi}. {We see that $\hat{\mathcal{C}}'$ and $\hat{\mathcal{T}}$ commute such that Eqs.~\eqref{eq:T1} and~\eqref{eq:T2} are identical statements.}

\end{listcases}

\subsection{$\mathcal{PT}$ conjugation}

Given the definitions of the parity and time-reversal operators, we have
\begin{subequations}
\begin{align}
\hat{\mathcal{P}}\hat{\mathcal{T}}\hat{\phi}_i(x)\hat{\mathcal{T}}^{-1}\hat{\mathcal{P}}^{-1}&=T_{ij}P_{jk}\check{\phi}_k(\mathcal{P}\mathcal{T}x)~,\\
\hat{\mathcal{P}}\hat{\mathcal{T}}\hat{\phi}_i^{\dag}(x)\hat{\mathcal{T}}^{-1}\hat{\mathcal{P}}^{-1}&=T_{ij}P_{jk}\check{\phi}_k^{\dag}(\mathcal{P}\mathcal{T}x)~;
\end{align}
\end{subequations}
and, taking $T_{ij}=\delta_{ij}$, it follows that
\begin{equation}
     \hat{\phi}^{\ddag}_i(x)=P_{ij}\check{\phi}_j^{\dag}(x)~,
 \end{equation}
since $\check{\phi}^{\mathsf{T}}(\mathcal{PT}x)=\check{\phi}^{\dag}(x)$. The $\mathcal{PT}$-symmetry of the Hamiltonians in Eqs.~\eqref{eq:hatHamiltonian} and~\eqref{eq:checkHamiltonian} is now readily confirmed. Note that, in Fock space, the requirement of $\mathcal{PT}$ symmetry is that $[\hat{H},\ddag]=0$, superseding the constraint of Hermiticity, i.e., $[\hat{H},\dag]=0$. This should be compared with the classical, and quantum-mechanical requirement, that $[\hat{H},\hat{\mathcal{P}}\hat{\mathcal{T}}]=0$. We can also easily check that $[\hat{\mathcal{C}}',\ddag]=0$, as required.

\section{Scalar-pseudoscalar mixing and oscillations}
 \label{sec:scalars}

We now illustrate the discussion in the previous sections by studying mixing and oscillations in the model with two spin-zero fields. As mentioned earlier, the Lagrangian (\ref{eq:scalLag}) and the corresponding Hamiltonian do not conserve parity. We therefore anticipate the possibility of scalar-pseudoscalar mixing and oscillations, but issues of interpretation arise (see Ref.~\cite{Ohlsson:2019noy, OZmixing}), as we now discuss in detail.

\subsection{Issues in Flavor Oscillations in the $\mathcal{PT}$-Symmetric Model}
\label{sec:PTmix}

In the mass eigenbasis (see Subsec.~\ref{sec:cdiscrete}), the classical equations of motion take the form
\be
\label{equamotXi}
\Box\xi_{\pm}+m_{\pm}^2\xi_{\pm}=0~,
\ee
which have the plane-wave solutions 
\be\label{planewave}
\xi_{\pm}=A_{\pm}e^{i[E_{\pm,\mathbf{p}} t-\mathbf{p}\cdot\mathbf{x}]}\qquad \text{with}\qquad E_{\pm,\mathbf{p}}=\sqrt{\mathbf{p}^2+m_\pm^2}~,
\ee
where $A_\pm$ are constants.

The single-particle flavour eigenstates can be written in terms of the mass eigenstates as follows:
\begin{equation}
     \ket{\check{\mathbf{p}},1(2),t}=\check{a}^{\dag}_{1(2),\mathbf{p}}(t)\ket{0}=N\left\{\eta\ket{\mathbf{p},+(-),t}-\left[1-\sqrt{1-\eta^2}\right]\ket{\mathbf{p},-(+),t}\right\}~.
 \end{equation}
As per the discussions of Secs.~\ref{sec:cdiscrete} and~\ref{sec:products}, the mass eigenstates are orthonormal with respect to the $\mathcal{C}'\mathcal{PT}$ inner product. The conjugate flavour state is
\begin{equation}
    \bra{\hat{\mathbf{p}},1(2),t}\equiv\bra{0}\hat{a}_{1(2),\mathbf{p}}(t)=N\left\{\eta\left(\ket{\mathbf{p},+(-),t}\right)^{\S}+\left[1-\sqrt{1-\eta^2}\right]\left(\ket{\mathbf{p},-(+),t}\right)^{\S}\right\}~,
\end{equation}
which has been expressed in terms of the $\mathcal{C}'\mathcal{PT}$-conjugate mass eigenstates by appealing to Eqs.~\eqref{eq:CprimePversusR} and~\eqref{eq:Rtrans2}.
The flavour and mass eigenstates obey the following orthonormality relations:
\begin{equation}
    \braket{\check{\mathbf{p}},i,t|\hat{\mathbf{p}}',j,t}=(2\pi)^3\delta_{ij}\delta^3(\mathbf{p}-\mathbf{p}')~,
\end{equation}
and
\begin{subequations}
\begin{align}
\left(\ket{\mathbf{p},\pm,t}\right)^{\S}\ket{\mathbf{p}',\pm,t}&=(2\pi)^3\delta^3(\mathbf{p}-\mathbf{p}')~,\\ \left(\ket{\mathbf{p},\pm,t}\right)^{\S}\ket{\mathbf{p}',\mp,t}&=0~.
\end{align}
\end{subequations}

Assuming for simplicity a localized initial state, the probability for the scalar with flavour $i$ at $t=0$ to transition to the pseudoscalar with flavour $j$ at $t>0$ is given naively by
\be
\Pi_{i\to j}(t)=\frac{1}{V}\int_{\mathbf{p}'}\braket{\check{\mathbf{p}},i,t|\hat{\mathbf{p}}',j,0}\braket{\check{\mathbf{p}}',j,0|\hat{\mathbf{p}},i,t}~,
\ee
where $V=(2\pi)^3\delta^3(\bm{0})$ is a three-volume. {We draw attention to the fact that this ``probability'' is not obtained from the usual squared modulus with respect to Hermitian conjugation --- were we to use this, we would find that the total probability is not conserved --- instead it involves the amplitude and its $\mathcal{C}'\mathcal{PT}$ conjugate.
A straightforward calculation then leads to
\begin{align}
\label{nonHproba}
\Pi_{i\to \slashed{i}}  (t)&=-\frac{\eta^2}{1-\eta^2}\sin^2\left(\frac{1}{2}(E_+(\mathbf{p})-E_-(\mathbf{p}))t\right)~.
\end{align}
Alarmingly, this ``probability" is negative, and the 
corresponding survival ``probability" is given by
\begin{equation}
    \Pi_{i\to i}  (t)=\frac{1}{V}\int_{\mathbf{p}'}\braket{\check{\mathbf{p}},i,t|\hat{\mathbf{p}}',i,0}\braket{\check{\mathbf{p}}',i,0|\hat{\mathbf{p}},i,t}=1+{\frac{\eta^2}{1-\eta^2}}\sin^2\left(\frac{1}{2}(E_+(\mathbf{p})-E_-(\mathbf{p}))t\right)~,
\end{equation}
which can be larger than unity. Notice, however, that $\Pi_{i\to i}+\Pi_{i\to\slashed{i}}=1$, such that the total ``probability'' is conserved.

It is interesting to note that the oscillation period obtained from the ``probability" \eqref{nonHproba} 
diverges at the exceptional points $\eta^2\to1$, where
\be
T=\frac{2\pi}{E_+(\mathbf{p})-E_-(\mathbf{p})}\simeq\frac{2\pi}{E_0(\mathbf{p})\sqrt{1-\eta^2}}
\qquad\text{with}\qquad E_0(\mathbf{p})\equiv\frac{m_1^2-m_2^2}{2\sqrt{\mathbf{p}^2+(m_1^2+m_2^2)/2}}~,
\ee
since the eigenmasses become degenerate. Another way to understand this limit is to consider the similarity transformation \eqref{eq:Rtrans} when $\eta\to\epsilon=\pm1$:
\be
\lim_{\eta\to\epsilon}\left\{\frac{R}{N}\right\}=\begin{pmatrix} \epsilon & 1 \\ 1 & \epsilon \end{pmatrix}~~~~\mbox{with}~~N\to\infty~.
\ee
We see that the {eigenstates} defined in Eq.~\eqref{equamotXi} are parallel in these limits.
Therefore, in addition to having infinite normalization, 
the similarity transformation is not invertible at the exceptional points, 
and one cannot define a map back to the flavour states.

It is illustrative to compare this oscillation ``probability" for the non-Hermitian theory to the corresponding probability for the Hermitian theory with the Lagrangian
\begin{equation}
    \hat{\mathcal{L}}_{\rm Herm}=\partial_{\nu}\hat{\phi}_i^{\dag}\partial^{\nu}\hat{\phi}_i-m_i^2\hat{\phi}_i^{\dag}\hat{\phi}_i-\mu^2\left(\hat{\phi}_1^{\dag}\hat{\phi}_2+\hat{\phi}_2^{\dag}\hat{\phi}_1\right)~,
\end{equation}
where $m_i^2$ and $\mu^2$ are positive real-valued squared mass parameters, and we assume $m_1^2>m_2^2$ as before. For this theory, the oscillation probability is
\be
\Pi_{i\to j}^{\rm Herm}(t)=\sin^2(2\alpha)\sin^2\left(\frac{1}{2}(E_+(\mathbf{p})-E_-(\mathbf{p}))~t\right)~,
\ee
where $\alpha$ is the mixing angle, which is given by
\be\label{Hmixingangle}
\sin^2(\alpha)=\frac{1}{2}-\frac{1}{2}\sqrt{1-\frac{4\mu^4}{(m_1^2-m_2^2)^2+4\mu^4}}~.
\ee
We see that the probability \eqref{nonHproba} has the same form as in the Hermitian case,
provided one makes the identification $\sin(2\alpha)=i\eta/\sqrt{1-\eta^2}$. With this identification, we have
\be
{\sin^2(\alpha)=\frac{1}{2}-\frac{1}{2}\sqrt{1+\frac{4\mu^4}{(m_1^2-m_2^2)^2-4\mu^4}}~,}
\ee
which is simply the analytic continuation $\mu^2\to i \mu^2$ of the Hermitian expression, and it is for this reason that the non-Hermitian transition ``probability'' is negative in the $\mathcal{PT}$-symmetric regime. As we show in the next Subsection, this problem has arisen because of an attempt to treat the flavour states as external states.

\subsection{Flavor Mixing in Scattering Matrix Elements}

The resolution of the above issues can be found by
recalling that experimental observables are scattering matrix elements,
for which we now give a simplified treatment. For this purpose,
we introduce two complex sources $J_A$ and $J_B$, coupled to the fields $\phi_1$ and $\phi_2$ as
\begin{equation}
    \label{eq:Lint}
    \mathcal{L}_{\rm int}=J_A\check{\phi}_1^{\dag}+J_A^{\dag}\hat{\phi}_1-J_B\check{\phi}_2^{\dag}+J_B^{\dag}\hat{\phi}_2~.
\end{equation}
We draw attention to the important fact that $\mathcal{L}_{\rm int}$ must be $\mathcal{PT}$ symmetric, giving rise to the relative minus sign between the last two terms~\cite{AEMS1, AEMS2}. The source terms are therefore not Hermitian.

The matrix element for the process $A\to B$ is given by
\begin{equation}
    i\mathcal{M}_{A\to B}=(2\pi)^4\delta^4(p_A-p_B)\Delta_{F,21}(q)~,
\end{equation}
with $q=p_A=p_B$, and the conjugate matrix element is
\begin{equation}
    -i\mathcal{M}_{A\to B}^{\mathcal{C}'\mathcal{PT}}=-(2\pi)^4\delta^4(p_A-p_B)\Delta_{D,12}(q)~.
\end{equation}
We note the overall sign, which stems from the relative sign in Eq.~\eqref{eq:Lint}.  The Feynman and Dyson propagators $\Delta_{F,ij}(q)$ and $\Delta_{D,ij}(q)$ are defined by
\begin{align*}
    \Delta_{F,ij}(x-y)&\equiv \braket{\mathrm{T}[\check{\phi}_i^{\dag}(x)\hat{\phi}_j(y)]}=\int\!\frac{{\rm d}^4q}{(2\pi)^4}e^{-iq.(x-y)}\Delta_{F,ij}(q)~,\\
    \Delta_{D,ij}(x-y)&\equiv \braket{\bar{\mathrm{T}}[\check{\phi}_i^{\dag}(x)\hat{\phi}_j(y)]}=\int\frac{{\rm d}^4q}{(2\pi)^4}e^{-iq.(x-y)}\Delta_{D,ij}(q)~,
\end{align*}
where $\mathrm{T}$ and $\bar{\mathrm{T}}$ denote time- and anti-time-ordering, respectively.  They can be calculated directly by inverting the non-Hermitian Klein-Gordon operator in momentum space, or by expressing the fields $\phi_1$ and $\phi_2$ in the mass eigenbasis, i.e.,
\begin{subequations}
\begin{align}
    \hat{\phi}_1=N\left[\eta\hat{\xi}_1-(1-\sqrt{1-\eta^2})\hat{\xi}_2\right]~,\\
    \check{\phi}^{\dag}_1=N\left[\eta\hat{\xi}^{\dag}_1+(1-\sqrt{1-\eta^2})\hat{\xi}^{\dag}_2\right]~,\\
    \hat{\phi}_2=N\left[\eta\hat{\xi}_2-(1-\sqrt{1-\eta^2})\hat{\xi}_1\right]~,\\
    \check{\phi}^{\dag}_2=N\left[\eta\hat{\xi}^{\dag}_2+(1-\sqrt{1-\eta^2})\hat{\xi}^{\dag}_1\right]~,
\end{align}
\end{subequations}
giving
\begin{subequations}
\begin{align}
    \Delta_{F,21}(q)&=N^2\eta(1-\sqrt{1-\eta^2})\left[\frac{i}{q^2-M_+^2+i\epsilon}-\frac{i}{q^2-M_-^2+i\epsilon}\right]\nonumber\\&=\frac{i\mu^2}{(q^2-M_+^2+i\epsilon)(q^2-M_-^2+i\epsilon)}~,\\
    \Delta_{D,12}(q)&=N^2\eta(1-\sqrt{1-\eta^2})\left[\frac{i}{q^2-M_+^2-i\epsilon}-\frac{i}{q^2-M_-^2-i\epsilon}\right]\nonumber\\&=\frac{i\mu^2}{(q^2-M_+^2-i\epsilon)(q^2-M_-^2-i\epsilon)}~.
\end{align}
\end{subequations}
Notice that, by virtue of the non-Hermiticity, we have
\begin{equation}
    \label{eq:FDrel}
    \Delta_{F,21}(q)=-\Delta^*_{D,12}(q)~.
\end{equation}
If the mass mixing were Hermitian, the Feynman and Dyson propagators would instead satisfy $\Delta^{(\mathrm{Herm})}_{F,21}(q)=\Delta^{(\mathrm{Herm})*}_{D,12}(q)$. The sign appearing in Eq.~\eqref{eq:FDrel} is due to the skew symmetry of the squared mass matrix.

We therefore have
\begin{subequations}
\begin{align}
    i\mathcal{M}_{A\to B}=(2\pi)^4\delta^4(p_A-p_B)N^2\eta\left(1-\sqrt{1-\eta^2}\right)\left[\frac{i}{s-M_+^2}-\frac{i}{s-M_-^2}\right]~,\\
    -i\mathcal{M}_{A\to B}^{\mathcal{C}'\mathcal{PT}}=-(2\pi)^4\delta^4(p_A-p_B)N^2\eta\left(1-\sqrt{1-\eta^2}\right)\left[\frac{i}{s-M_+^2}-\frac{i}{s-M_-^2}\right]~.
\end{align}
\end{subequations}
where $s=p_A^2=p_B^2$ is the usual Mandelstam variable and we have suppressed the pole prescription in the propagators. The squared matrix element is then
\begin{align}
    \mathcal{M}_{A\to B}^{\mathcal{C}'\mathcal{PT}}\mathcal{M}_{A\to B}&=VT(2\pi)^4\delta^4(p_A-p_B)N^4\eta^2\left(1-\sqrt{1-\eta^2}\right)^2\left[\frac{1}{s-M_+^2}-\frac{1}{s-M_-^2}\right]^2\nonumber\\&=\frac{1}{4}VT(2\pi)^4\delta^4(p_A-p_B)\frac{\eta^2}{1-\eta^2}\left[\frac{1}{s-M_+^2}-\frac{1}{s-M_-^2}\right]^2~,
    \label{trick}
\end{align}
where $VT\equiv(2\pi)^4\delta^4(0)$ is a four-volume factor.

We observe that this result is positive for $\eta^2<1$. However, it would seem naively that there is an issue with perturbative unitarity in the limit $\eta^2 \to 1$, due to the factor of $1/(1 - \eta^2)$ in (\ref{trick}),
but this is not the case, since
\begin{equation}
    \left[\frac{1}{s-M_+^2}-\frac{1}{s-M_-^2}\right]^2=\frac{\left(M_+^2-M_-^2\right)^2}{\left(s-M_+^2\right)^2\left(s-M_-^2\right)^2}=\left(1-\eta^2\right)\frac{\left(m_1^2-m_2^2\right)^2}{\left(s-M_+^2\right)^2\left(s-M_-^2\right)^2}
\end{equation}
is proportional to $1-\eta^2$. The final expression for the squared matrix element is
\begin{equation}
    \mathcal{M}_{A\to B}^{\mathcal{C}'\mathcal{PT}}\mathcal{M}_{A\to B}=VT(2\pi)^4\delta^4(p_A-p_B)\frac{\mu^4}{\left(s-M_+^2\right)^2\left(s-M_-^2\right)^2}~,
\end{equation}
which is positive, vanishes in the limit $\mu\to 0$ (as it should), and remains real and perturbatively valid all the way up to the exceptional point $\eta^2=1$.

The matrix element for the corresponding flavour-conserving process $A\to A$ is
\begin{equation}
    i\mathcal{M}_{A\to A}=(2\pi)\delta^4(p_A-p_A')\Delta_{F,11}(q)~,
\end{equation}
with $q=p_a=p_A'$, and the conjugate matrix element is
\begin{equation}
    -i\mathcal{M}_{A\to A}^{\mathcal{C}'\mathcal{PT
}}=(2\pi)\delta^4(p_A-p_A')\Delta_{D,11}(q)~,
\end{equation}
where
\begin{equation}
\Delta_{D,11}(q)=\Delta_{F,11}^*(q)~,
\end{equation}
with
\begin{equation}
    \Delta_{F,11}(q)=N^2\Bigg[\frac{i\eta^2}{q^2-M_+^2+i \epsilon}-\frac{i(1-\sqrt{1-\eta^2})^2}{q^2-M_-^2+i \epsilon}\Bigg]=\frac{i(q^2-m_2^2)}{(q^2-M_+^2+i \epsilon)(q^2-M_-^2+i \epsilon)}~.
\end{equation}
We therefore obtain
\begin{equation}
    \mathcal{M}_{A\to A}^{\mathcal{C}'\mathcal{PT
}}\mathcal{M}_{A\to A}=VT(2\pi)^4\delta^4(p_A-p_A')\frac{(s-m_2^2)^2}{(s-M_+^2)^2(s-M_-^2)^2}~,
\end{equation}
which is again real, positive and physically meaningful for all $0\leq\eta^2\leq 1$.

It is clear from these results that there is a subtlety arising from the factorisation of the source-to-source probability into production, oscillation and detection probabilities. This offers a resolution of the problematic behaviour in the naive calculation of the oscillation probability presented in Subsec.~\ref{sec:PTmix}. Comforted by the example presented in this Subsection, we leave for future work the further detailed study of this point.

\section{Conclusions}
\label{sec:conx}

We have addressed in this paper some basic issues in the formulation of non-Hermitian 
bosonic quantum field theories, discussing in particular the
treatment of discrete symmetries and the definition of the inner product in Fock space. We have focused on  $\mathcal{PT}$-symmetric non-Hermitian theories,
commenting also on features at the exceptional points at the boundary between theories with  
$\mathcal{PT}$ symmetry and those in which it is broken.

As we have discussed, there is ambiguity in the formulation of the inner product in a $\mathcal{PT}$-symmetric 
theory. In this case, the conventional Dirac inner product  $(\ket{\alpha})^{\dag}\ket{\beta}=\braket{\alpha^*|\beta}$ 
is not positive definite {for the mass eigenstates}, and is therefore deprecated, and the same is true of of the $\mathcal{PT}$
inner product $(\ket{\alpha})^{\ddag}\ket{\beta}=\braket{\alpha^{\mathcal{PT}}|\beta}$, where
$\ddag\equiv \mathcal{PT}\circ \mathsf{T}$ with $\mathsf{T}$ denoting transposition. The appropriate positive-definite norm {for the mass eigenstates} is defined via
$\mathcal{C'PT}$ conjugation: $(\ket{\alpha})^{\S}\ket{\beta}=\braket{\alpha^{\mathcal{C'PT}}|\beta}$,
where $\S\equiv \mathcal{C'PT}\circ \mathsf{T}$, where the $\mathcal{C'}$ operator was defined in Subsec.~\ref{sec:Cprime}. As was explained there, the $\mathcal{C}'$ 
transformation in a $\mathcal{PT}$-symmetric quantum field theory
cannot be identified with charge conjugation.

We have formulated in Subsec.~\ref{sec:similarity} a suitable similarity transformation between a $\mathcal{PT}$-symmetric non-Hermitian theory with two flavours  of spin-zero fields and its Hermitian counterpart. 
The equivalence between the non-interacting $\mathcal{PT}$-symmetric and 
Hermitian theories cannot, in general, be carried over to interacting theories with the same similarity transformation.
Appendix~\ref{sec:alternativesim} contrasts the similarity transformation we propose with the previous literature.

As an illustration of this Fock-space discussion, 
we have considered mixing and oscillations in this specific model with two boson flavours, which is free apart from
non-Hermitian $\mathcal{PT}$-symmetric mixing terms. The unmixed bosons are taken to
be a scalar and a pseudoscalar, which mix via a non-Hermitian bilinear term. We have
shown that the resulting mass eigenvectors are not orthogonal with respect to the 
Dirac inner product, but are orthogonal with positive norm when the 
$\mathcal{C'PT}$ inner product is used. {We have emphasized that the parity operator
in this two-boson model does not commute with the Hamiltonian, leading to the appearance of
scalar-pseudoscalar mixing and flavour oscillations, which we have studied in Sec.~\ref{sec:scalars}.
These} are of similar form to the mixing
between bosons in a Hermitian theory, respecting unitarity but presenting issues of interpretation, which we show can be resolved by considering physical scattering matrix elements, wherein flavour states only appear internally.

The analysis in this paper has clarified the description of
$\mathcal{PT}$-symmetric non-Hermitian bosonic quantum field theories, and provides a
framework for formulating them off-shell. Many of the features discussed here are expected to carry over to $\mathcal{PT}$-symmetric non-Hermitian field theories of fermions{~\cite{nonHFermions}}, as we  shall discuss in a following paper. This programme constitutes an important step towards addressing deeper issues in field theory such as quantum loop corrections and renormalization, to which we also plan to return in future work.

\section*{Acknowledgements}

PM thanks Maxim Chernodub for helpful comments on the manuscript. The work of JA and JE was supported by the United Kingdom STFC Grant No.\ ST/P000258/1, 
and that of JE also by the Estonian Research Council via a Mobilitas Pluss grant. 
The work of PM was supported by a Leverhulme Trust Research Leadership Award (Grant No.~RL-2016-028) and a Nottingham Research Fellowship from the University of Nottingham.

\appendix
\setcounter{equation}{0}
\section{An alternative similarity transformation}
\label{sec:alternativesim}

A different similarity transformation~\cite{Mannheim} has previously been applied to the boson model considered in this work. In this Appendix, we review it for completeness, and make a comparison with the transformation detailed in Subsec.~\ref{sec:similarity}.

The Hamiltonian $\hat{H}$ of the two-flavour scalar theory can also be mapped to a Hermitian one $\hat{h}_{\mathcal{S}}$ (and similarly for the Lagrangian) via the similarity transformation{~\cite{Mannheim}}
\begin{equation}
    \hat{h}_{\mathcal{S}}=\hat{\mathcal{S}}\hat{H}\hat{\mathcal{S}}^{-1}~,
\end{equation}
with
\begin{equation}
    \label{eq:noetasimilarity}
    \hat{\mathcal{S}}=\exp\left[\frac{\pi}{2}\int_{\mathbf{x}}\left(\hat{\pi}_2(t,\mathbf{x})\hat{\phi}_2(t,\mathbf{x})+\hat{\phi}_2^{\dag}(t,\mathbf{x})\hat{\pi}_2^{\dag}(t,\mathbf{x})\right)\right]~.
\end{equation}
Here, we have written the operator $\hat{\mathcal{S}}$ in a manifestly Hermitian form. We note, however, that the similarity transformation is defined only up to a constant complex phase, such that one is free to reorder the operators in the exponent by making use of the canonical equal-time commutation relations. We note that, unlike the similarity transformation we propose in the main text, the transformation \eqref{eq:noetasimilarity} does not depend on the non-Hermitian parameter $\eta$.

The similarity transformation {\eqref{eq:noetasimilarity}} has the following action on the field operators:
\begin{subequations}
\label{eq:noetafieldsimilarity}
\begin{align}
    \hat{\mathcal{S}}\hat{\phi}_2(t,\mathbf{x})\hat{\mathcal{S}}^{-1}=-i\hat{\phi}_2(t,\mathbf{x})~,\\ \hat{\mathcal{S}}\hat{\phi}_2^{\dag}(t,\mathbf{x})\hat{\mathcal{S}}^{-1}=-i\hat{\phi}_2^{\dag}(t,\mathbf{x})~,
\end{align}
\end{subequations}
and the transformed {version of the Lagrangian \eqref{eq:scalLaghat}} for the free scalar theory is therefore
\begin{equation}
    \hat{\mathcal{L}}_{\mathcal{S}}=\partial_{\nu}\hat{\phi}^{\dag}_1\partial^{\nu}\hat{\phi}_1-\partial_{\nu}\hat{\phi}^{\dag}_2\partial^{\nu}\hat{\phi}_2-m_1^2\hat{\phi}_1^{\dag}\hat{\phi}_1+m_2^2\hat{\phi}_2^{\dag}\hat{\phi}_2-i\mu^2(\hat{\phi}_1^{\dag}\hat{\phi}_2-\hat{\phi}_2^{\dag}\hat{\phi}_1)~.
\end{equation}
While this Lagrangian is Hermitian, we draw attention to the opposite relative signs of the kinetic and mass terms for the fields $\hat{\phi}_{1,2}$, which imply that $\hat{\phi}_{2}$
is a negative-norm ghost and is tachyonic. One should therefore suspect that the similarity transformation in Eq.~\eqref{eq:noetasimilarity} is not directly related to the $\hat{\mathcal{C}}'$ operator needed to construct a positive norm for these states. Moreover, one can readily confirm that this similarity transformation, unlike the one defined in Subsec.~\ref{sec:similarity}, does not leave the Fock vacuum invariant.

The latter issue is most easily illustrated by decoupling the two flavours, i.e., taking the Hermitian limit $\eta\to 0$. The plane-wave decomposition of the field $\hat{\phi}_2$ then takes a simple form, and we can immediately write
\begin{equation}
    \hat{\mathcal{S}}\big|_{\eta\to 0}\equiv \hat{\mathcal{S}}_0=\exp\left[i\frac{\pi}{2}\int_{\mathbf{p}}\left(\hat{a}_{2,\mathbf{p}}^{\dag}(0)\hat{c}^{\dag}_{2,-\mathbf{p}}(0)e^{2iE_{2,\mathbf{p}}t}-\hat{c}_{2,\mathbf{p}}(0)\hat{a}_{2,-\mathbf{p}}(0)e^{-2iE_{2,\mathbf{p}}t}\right)\right]~.
\end{equation}
The creation and annihilation operators transform as follows:
\begin{subequations}
 \begin{align}
     \hat{\mathcal{S}}_0\hat{a}_{2,\mathbf{q}}(0)\hat{\mathcal{S}}_0^{-1}&=-ie^{2iE_{2,\mathbf{q}}t}\hat{c}^{\dag}_{2,-\mathbf{q}}(0)~,\\
     \hat{\mathcal{S}}_0\hat{a}_{2,\mathbf{q}}^{\dag}(0)\hat{\mathcal{S}}_0^{-1}&=-ie^{-2iE_{2,\mathbf{q}}t}\hat{c}_{2,-\mathbf{q}}(0)~,\\
     \hat{\mathcal{S}}_0\hat{c}_{2,\mathbf{q}}(0)\hat{\mathcal{S}}_0^{-1}&=-ie^{2iE_{2,\mathbf{q}}t}\hat{a}^{\dag}_{2,-\mathbf{q}}(0)~,\\
     \hat{\mathcal{S}}_0\hat{c}_{2,\mathbf{q}}^{\dag}(0)\hat{\mathcal{S}}_0^{-1}&=-ie^{-2iE_{\mathbf{q}}t}\hat{a}_{2,-\mathbf{q}}(0)~,
 \end{align}
\end{subequations}
which are consistent with the transformations of the fields in Eq.~\eqref{eq:noetafieldsimilarity}. This transformation would lead to the following candidate $\hat{\mathcal{C}}'$ operator:
\begin{equation}
\label{eq:candidate}
    \hat{\mathcal{C}}_?'=\exp\left[i\pi\int_{\mathbf{p}}\left(\hat{a}_{2,\mathbf{p}}^{\dag}(0)\hat{c}^{\dag}_{2,-\mathbf{p}}(0)e^{2iE_{\mathbf{p}}t}-\hat{c}_{2,\mathbf{p}}(0)\hat{a}_{2,-\mathbf{p}}(0)e^{-2iE_{2,\mathbf{p}}t}\right)\right]\hat{\mathcal{P}}~.
\end{equation}
However, we see immediately that this operator does not leave the Fock vacuum  invariant. Instead, it is transformed to an infinite series of time-dependent multiparticle states:
\begin{align}
    \hat{\mathcal{C}}'_?\ket{0}&=\left(1+\frac{\pi^2}{2!}+\cdots\right)\left(\ket{0}+i\pi\int_{\mathbf{p}}\ket{\mathbf{p},2,t;\bar{\mathbf{p}},2,t}\right.\nonumber\\&\qquad\left.+\frac{(i\pi)^2}{2!}\int_{\mathbf{p},\mathbf{q}}\ket{\mathbf{p},2,t;\bar{\mathbf{p}},2,t;\mathbf{q},2,t;\bar{\mathbf{q}},2,t}+\cdots\right)~,
\end{align}
wherein antiparticle states are indicated by a bar over the three-momentum with, e.g., $\bar{\mathbf{p}}=-\mathbf{p}$.

\section{Some useful expressions}
\label{sec:useful}

In this Appendix, we collect useful expressions for the various mass and flavour states. The ket states are as follows:
\begin{subequations}
\begin{align}
    \ket{\check{\mathbf{p}},1,t}=N\left\{\eta\ket{\mathbf{p},+,t}-(1-\sqrt{1-\eta^2})\ket{\mathbf{p},-,t}\right\}~,\\
    \ket{\check{\mathbf{p}},2,t}=N\left\{\eta\ket{\mathbf{p},-,t}-(1-\sqrt{1-\eta^2})\ket{\mathbf{p},+,t}\right\}~,\\
    \ket{\mathbf{p},+,0}=N\left\{\eta\ket{\mathbf{p},1,0}+(1-\sqrt{1-\eta^2})\ket{\mathbf{p},2,0}\right\}~,\\
    \ket{\mathbf{p},-,0}=N\left\{\eta\ket{\mathbf{p},2,0}+(1-\sqrt{1-\eta^2})\ket{\mathbf{p},1,0}\right\}~.
\end{align}
\end{subequations}
We also have that
\begin{subequations}
\begin{align}
    \ket{\mathbf{p},+,t}=Ne^{iE_+t}\left\{\eta\ket{\mathbf{p},1,0}+(1-\sqrt{1-\eta^2})\ket{\mathbf{p},2,0}\right\}~,\\
    \ket{\mathbf{p},-,t}=Ne^{iE_-t}\left\{\eta\ket{\mathbf{p},2,0}+(1-\sqrt{1-\eta^2})\ket{\mathbf{p},1,0}\right\}~.
\end{align}
\end{subequations}
The distinction between checked and hatted operators is not needed for the flavour eigenstates at the initial time or mass eigenstates for all times. The conjugate states are
\begin{subequations}
\begin{align}
    \label{eq:t_dep_1}
    \bra{\hat{\mathbf{p}},1,t}=N\left\{\eta\bra{\mathbf{p},+,t}+(1-\sqrt{1-\eta^2})\bra{\mathbf{p},-,t}\right\}~,\\
    \bra{\hat{\mathbf{p}},2,t}=N\left\{\eta\bra{\mathbf{p},-,t}+(1-\sqrt{1-\eta^2})\bra{\mathbf{p},+,t}\right\}~,\\
    \bra{\mathbf{p},+,0}\hat{\mathcal{C}}'\hat{\mathcal{P}}\hat{\mathcal{T}}=N\left\{\eta\bra{\mathbf{p},1,0}-(1-\sqrt{1-\eta^2})\bra{\mathbf{p},2,0}\right\}~,\\
    \bra{\mathbf{p},-,0}\hat{\mathcal{C}}'\hat{\mathcal{P}}\hat{\mathcal{T}}=N\left\{\eta\bra{\mathbf{p},2,0}-(1-\sqrt{1-\eta^2})\bra{\mathbf{p},1,0}\right\}~,
\end{align}
\end{subequations}
with
\begin{subequations}
\begin{align}
    \bra{\mathbf{p},+,t}\hat{\mathcal{C}}'\hat{\mathcal{P}}\hat{\mathcal{T}}=Ne^{-iE_+t}\left\{\eta\bra{\mathbf{p},1,0}-(1-\sqrt{1-\eta^2})\bra{\mathbf{p},2,0}\right\}~,\\
    \bra{\mathbf{p},-,t}\hat{\mathcal{C}}'\hat{\mathcal{P}}\hat{\mathcal{T}}=Ne^{-iE_-t}\left\{\eta\bra{\mathbf{p},2,0}-(1-\sqrt{1-\eta^2})\bra{\mathbf{p},1,0}\right\}~.
\end{align}
\end{subequations}
The orthogonality relations for the mass eigenstates are therefore
\begin{subequations}
\begin{align}
\bra{\mathbf{p},\pm,t}\hat{\mathcal{C}}'\hat{\mathcal{P}}\hat{\mathcal{T}}\ket{\mathbf{p}',\pm,t}&=(2\pi)^3\delta^3(\mathbf{p}-\mathbf{p}')~,\\
\bra{\mathbf{p},\pm,t}\hat{\mathcal{C}}'\hat{\mathcal{P}}\hat{\mathcal{T}}\ket{\mathbf{p}',\mp,t}&=0~,
\end{align}
\end{subequations}
since
\begin{subequations}
\begin{align}
    \braket{\mathbf{p},i,0|\mathbf{p}',j,0}=(2\pi)^3\delta_{ij}\delta^3(\mathbf{p}-\mathbf{p}')
\end{align}
\end{subequations}
by virtue of the algebra in Sec.~\ref{sec:quant}. Moreover, we have that
\begin{align}
    \ket{\check{\mathbf{p}},1,t}&=\left(e^{iE^{\mathsf{T}}t}\right)_{1i}\ket{\check{\mathbf{p}},i,0}\nonumber\\&=\left(e^{iE^{\mathsf{T}}t}\right)_{1i}N\begin{pmatrix} \eta \ket{\mathbf{p},+,0} - (1-\sqrt{1-\eta^2})\ket{\mathbf{p},-,0} \\ \eta \ket{\mathbf{p},-,0} - (1-\sqrt{1-\eta^2})\ket{\mathbf{p},+,0}\end{pmatrix}_i
    \nonumber\\&=\left(e^{iE^{\mathsf{T}}t}\right)_{1i}R^{-1}_{ij}\begin{pmatrix} \ket{\mathbf{p},+,0}\\ \ket{\mathbf{p},-,0}\end{pmatrix}_j\nonumber\\&=R^{-1}_{1i}\left(e^{iE^{\rm diag}t}\right)_{ij}R_{jk}R^{-1}_{k\ell}\begin{pmatrix} \ket{\mathbf{p},+,0}\\ \ket{\mathbf{p},-,0}\end{pmatrix}_{\ell}\nonumber\\&=R_{1i}^{-1}\begin{pmatrix} e^{iE_+t}\ket{\mathbf{p},+,0}\\ e^{iE_-t}\ket{\mathbf{p},-,0}\end{pmatrix}_{i}~,
\end{align}
which is consistent with Eq.~\eqref{eq:t_dep_1}.

\end{document}